\documentclass[prx,superscriptaddress,amsfonts,amssymb,amsmath,floats,twocolumn,aps,footinbib,shownopacs]{revtex4}
\usepackage[pdftex]{graphicx}
\usepackage{subfigure}
\usepackage{dsfont}
\usepackage{lipsum}
\usepackage{dcolumn}
\usepackage{bm}
\usepackage{color}
\usepackage{physics}
\usepackage{multirow}
\usepackage[pdftex,colorlinks=true, linkcolor=blue,citecolor=blue,filecolor=blue]{hyperref}
\usepackage{array}
\newcommand{\PreserveBackslash}[1]{\let\temp=\\#1\let\\=\temp}
\newcolumntype{C}[1]{>{\PreserveBackslash\centering}p{#1}}
\newcolumntype{R}[1]{>{\PreserveBackslash\raggedleft}p{#1}}
\newcolumntype{L}[1]{>{\PreserveBackslash\raggedright}p{#1}}
\usepackage{pifont}% http://ctan.org/pkg/pifont
\newcommand{\cmark}{\ding{51}}%
\newcommand{\xmark}{\ding{55}}%

\begin{document}
\title{Theory of nematic charge orders in kagome metals}

\author{Francesco Grandi} 
\affiliation{Institut f\"ur Theorie der Statistischen Physik, RWTH Aachen University, 52056 Aachen, Germany}
\author{Armando Consiglio}
\affiliation{Institut f\"ur Theoretische Physik und Astrophysik and W\"urzburg-Dresden Cluster of Excellence ct.qmat, Universit\"at W\"urzburg, 97074 W\"urzburg, Germany}
\author{Michael A. Sentef}
\affiliation{H H Wills Physics Laboratory, University of Bristol, Bristol BS8 1TL, United Kingdom}
\affiliation{Max Planck Institute for the Structure and Dynamics of Matter, Center for Free-Electron Laser Science (CFEL), Luruper Chaussee 149, 22761 Hamburg, Germany}
\author{Ronny Thomale}
\affiliation{Institut f\"ur Theoretische Physik und Astrophysik and W\"urzburg-Dresden Cluster of Excellence ct.qmat, Universit\"at W\"urzburg, 97074 W\"urzburg, Germany} 
\affiliation{Department of Physics and Quantum Centers in Diamond and Emerging Materials (QuCenDiEM) group, Indian Institute of Technology Madras, Chennai 600036, India}
\author{Dante M. Kennes}
\affiliation{Institut f\"ur Theorie der Statistischen Physik, RWTH Aachen University, 52056 Aachen, Germany}
\affiliation{JARA-Fundamentals of Future Information Technology, 52056 Aachen, Germany}
\affiliation{Max Planck Institute for the Structure and Dynamics of Matter, Center for Free-Electron Laser Science (CFEL), Luruper Chaussee 149, 22761 Hamburg, Germany}

\begin{abstract}
%FG
Kagome metals $A$V$_3$Sb$_5$ ($A=$K, Rb, Cs) exhibit an exotic charge order (CO), involving three order parameters, with broken translation and time-reversal symmetries compatible with the presence of orbital currents. The properties of this phase are still intensely debated, and it is unclear if the origin of the CO is mainly due to electron-electron or electron-phonon interactions. Most of the experimental studies confirm the nematicity of this state, a feature that might be enhanced by electronic correlations. However, it is still unclear whether the nematic CO becomes stable at a temperature equal to ($T_{\text{nem}} = T_\text{C}$) or lower than ($T_{\text{nem}} < T_\text{C}$) the one of the CO itself. Here, we systematically characterize several CO configurations, some proposed for the new member of the family ScV$_6$Sn$_6$, by combining phenomenological Ginzburg-Landau theories, valid irrespective of the specific ordering mechanism, with mean-field analysis. We find a few configurations for the CO that are in agreement with most of the experimental findings to date and that are described by different Ginzburg-Landau potentials. We propose to use resonant ultrasound spectroscopy to experimentally characterize the order parameters of the CO, such as the number of their components and their relative amplitude, and provide an analysis of the corresponding elastic tensors. This might help understand which mean-field configuration found in our study is the most representative for describing the CO state of kagome metals, and it can provide information regarding the nematicity onset temperature $T_\text{nem}$ with respect to $T_\text{C}$.
\end{abstract} 
\maketitle

%%%%%%%%%%%%%%%%%%%%%%%%%%%%%%%%%%%%%%%%%%%%%%% INTRO
\section{Introduction}
The interplay between electronic correlations and non-trivial band features has become a major topic in condensed matter physics, which is particularly prominently reflected in kagome metals \cite{Neupert2022_NatPhys,Jiang2023_NSR}. Indeed, the band structure of the kagome lattice hosts Dirac cones, van Hove singularities and a flat band, which, in the presence of strong electron-electron repulsion, might support several kinds of orderings, including non-trivial topological states. The compounds $A$V$_3$Sb$_5$ ($A=$K, Rb, Cs) form a hexagonal lattice with space group P$6/$mmm, and show a layered kagome lattice formed by the vanadium atoms. The V-$3$d orbitals contribute to most of the states at the Fermi level and they are responsible for the presence of several van Hove singularities close to zero energy \cite{Ortiz2019_PRM}. The low-energy band structure of these systems can be conveniently described with orbitals belonging to a single kagome layer, suggesting the almost two-dimensional nature of their electronic properties \cite{Wu2021_PRL}. By lowering the temperature of these compounds below $T_\text{C} \sim 90$K, kagome metals display the onset of a $2 \times 2$ in-plane charge-order (CO), characterized by the presence of three ordering peaks ($3$Q) seen in x-ray diffraction and scanning tunneling microscopy measurements \cite{Ortiz2020_PRL,Zhao2021_Nat,Tan2021_PRL}, pointing to the presence of three independent order parameters $\Delta_1$, $\Delta_2$ and $\Delta_3$ \cite{Kiesel2013_PRL,Wang2013_PRB,Venderbos2016_PRB}. This state might have additional $\times 2$ or $\times 4$ out-of-plane components \cite{Jiang2021_Nat_Mat,Li2022_Nat_Phys,Li2021_PRX,Liang2021_PRX,Hu2022_PRB} which can even coexist \cite{Xiao2023_PRR}, but recent resonant elastic x-ray scattering measurements performed on CsV$_3$Sb$_5$ suggest the presence of two COs characterized by $2 \times 2 \times 1$ and $2 \times 2 \times 2$ unit cell, with the former involving the vanadium $3$d-orbitals and the latter involving the antimony $5$p-orbitals \cite{Li2022_NatComm}, not in disagreement with simulations that underline the role of the Sb-related bands in the formation of the three-dimensional ordered state \cite{Tsirlin2022_SciPost}. Despite the fact that, by simple electron counting, one would expect a magnetic V$^{4+}$ ion per formula unit in the ionic limit, the phase diagram of these compounds does not show features of emergent magnetism \cite{Kenney2021_JoPCondMat}.

The bilayer material ScV$_6$Sn$_6$, similarly to other kagome metals, has kagome nets formed by vanadium atoms. The onset of non-magnetic CO has recently been found in this compound at the transition temperature $\sim 92$K \cite{Arachchige2022_PRL}, close to $T_\text{C}$ measured for $A$V$_3$Sb$_5$ ($A=$K, Rb, Cs). 
%FG
This CO shows the features of $3$Q ordering, i.e., the presence of three independent order parameters, but it is characterized by a $\sqrt{3} \times \sqrt{3} (\times 3)$ reconstruction \cite{Arachchige2022_PRL,Hu2022_arXiv,Cheng2023_arXiv}. 
%This CO shows the features of $3$Q ordering, but it is characterized by a $\sqrt{3} \times \sqrt{3}$ ($\times 3$) reconstruction \cite{Arachchige2022_PRL,Hu2022_arXiv}. 
Similar translational symmetry breaking might also be relevant for CsV$_3$Sb$_5$ \cite{Chen2021_Nat} and for the ground state of the extended Hubbard model on the kagome lattice \cite{Ferrari2022_PRB}. Since ScV$_6$Sn$_6$ belongs to the large family of hexagonal HfFe$_6$Ge$_6$-type compounds, it holds the promise of broad tuning opportunities.

Several experiments, including c-axis resistivity \cite{Xiang2021_NatComm}, nuclear magnetic resonance (NMR) \cite{Nie2022_Nat}, muon-spin relaxation/rotation ($\mu$SR) \cite{Yu2021_arXiv}, polarization resolved Raman spectroscopy \cite{Wu2022_PRB} and optical polarization rotation \cite{Wu2022_PRB2} have confirmed that CO in $A$V$_3$Sb$_5$ ($A=$K, Rb, Cs) has only twofold (C$_2$) symmetry, which is believed to be connected to electronic nematicity. This is also related to the chiral nature of the state, i.e., the broken in-plane mirror symmetry, which has been observed by measuring the electronic magnetochiral anisotropy \cite{Guo2022_Nat}. However, it is still unclear at which temperature electronic nematicity sets in. Several experiments suggest a reduction of the rotational symmetry of the system from C$_6$ to C$_2$ at the transition temperature of the CO, as confirmed by micron-scale spatially-resolved angle-resolved photoelectron spectroscopy (ARPES) on KV$_3$Sb$_5$ \cite{Jiang2022_arXiv} and by scanning birefringence microscopy on all the three compounds $A$V$_3$Sb$_5$ ($A=$K, Rb, Cs) \cite{Xu2022_NatPhys}. 
%FG
This reduced rotational symmetry might be due to a $\pi$-shift between the $2 \times 2$ CO in two consecutive kagome layers, implying that at least two of the order parameters might also have an out-of-plane component \cite{Christensen2021_PRB,Ratcliff2021_PRM}. Within this interpretation, a single kagome layer would retain the original sixfold rotational symmetry of the lattice even beyond the onset of the CO, i.e., the in-plane components of the order parameters satisfy the relation $| \Delta_1 | = | \Delta_2 | = | \Delta_3 |$. However, the stacking of different layers reduces the symmetry of the system to C$_2$. This way of lowering the rotational symmetry can be regarded as ``weak'' nematicity, as opposed to the ``strong'' one which is found at much lower temperatures than the CO, $T_\text{nem} \sim 30$K \cite{Zheng2022_Nat,Nie2022_Nat}, where the system explicitly breaks the sixfold rotational symmetry of each kagome layer. Thus, for a two-dimensional kagome lattice, the ``weak'' nematic regime would be characterized by $| \Delta_1 | = | \Delta_2 | = | \Delta_3 |$, while the ``strong'' nematic regime has $| \Delta_1 | = | \Delta_3 | \neq | \Delta_2 |$ (one component has to be different from the other two). The onset of the nematic CO at the transition temperature for the translational symmetry breaking would imply the presence of a nematic metal above $T_\text{C}$ (at the transition point, the ordered state must inherit the point group symmetries of the high-temperature phase \cite{Birman1966_PRL}), which might be related to anisotropic fluctuations of the order parameters \cite{Fernandes2011_PRL,Bouhmer2022_NatPhys}.

Other measurements have confirmed that the CO observed in kagome metals breaks time-reversal symmetry (TRS), suggesting the three order parameters $\Delta_j$ to be complex. The breaking of TRS is confirmed by $\mu$SR experiments \cite{Mielke2022_Nat,Khasanov2022_PRR,Yu2021_arXiv}, by the detection of a giant anomalous Hall effect \cite{Yang2020_SciAdv,Yu2021_PRB}, and by magneto-optical Kerr measurements \cite{Wu2022_PRB2,Hu2022_arXiv2,Xu2022_NatPhys}, even if more recent analyses question these observations \cite{Li2022_PRB,Saykin2022_arXiv,Wang2023_arXiv}. The absence of magnetic ordering together with broken TRS suggested an orbital current state as a possible candidate for this phase, akin to the ones described by the Haldane model on the honeycomb lattice or by the Varma model on the Lieb lattice \cite{Haldane1988_PRL,Varma1997_PRB}.

Understanding the leading mechanism for the onset of the CO is interesting \textit{per se}. However, this becomes even more relevant considering that, by lowering the temperature of $A$V$_3$Sb$_5$ ($A=$K, Rb, Cs) below $\sim 2$K \cite{Ortiz2020_PRL,Ortiz2021_PRM,Yin2021_ChinPhysLett}, these systems show superconducting (SC) features. The SC state might inherit the properties of the higher-temperature phase \cite{Guguchia2023_NatComm} and might be stabilized by charge fluctuations \cite{Tazai2022_SciAdv}. Yet the experimental evidence collected so far has not settled the debate about the leading mechanism causing the SC instability. The critical temperatures for the SC transition estimated from the electron-phonon coupling strength are lower than the experimental values, suggesting an important role of electronic correlations \cite{Tan2021_PRL,Wu2021_PRL}. Furthermore, the absence of the Kohn anomaly in inelastic x-ray scattering measurements at the onset of the CO indicates small electron-phonon interactions (EPI) in these materials \cite{Li2021_PRX}, even if the weak first-order character of the transition might explain the lack of this feature \cite{Mu2021_ChinPhysSoc,Song2022_SciChinPhys,Luo2022_npj,Li2022_NatComm}. Instead, recent ARPES \cite{Luo2022_NatComm,Zhong2022_arXiv} and Raman scattering (RS) \cite{Liu2022_NatComm} measurements underline the relevant role of the momentum dependence and of the strong local EPI \cite{Rossnagel2011_JPhys}, respectively, in the stabilization of the CO, in contrast to previous findings. Moreover, time-resolved ARPES signals registered during the melting of the CO were interpreted to confirm the critical role of phonons in the stabilization of the CO state \cite{Azoury2023_arXiv}.

The theoretical analysis that has been conducted so far for this class of compounds does not unambiguously solve the conundrum of the origin of the CO in the kagome metals. Indeed, several proposals have been advanced in the literature, some of them suggesting an electronic mechanism based on a generalized Peierls instability with wave vectors that correspond to the three inequivalent momenta (3Q) that connect the M-points of the Brillouin zone \cite{Tan2021_PRL}, and some pointing at the crucial role of the saddle point van Hove singularities in the electronic band structure at the Fermi level to drive the CO \cite{Rice1975_PRL,Zhou2021_PRB}, underlining that the Peierls mechanism cannot always properly describe the occurrence of the ordering instability in real compounds \cite{Johannes2008_PRB}. Other analyses suggest a prominent role of the EPI. However, it is not clear whether the $k$-dependence \cite{Si2022_PRB} or the local (Jahn-Teller) character of the EPI \cite{Wang2022_PRB,Ptok2022_PRB} is more important. The discrepancy among these interpretations becomes even more baffling considering that all of them are obtained by performing simulations based on density functional theory (DFT) with exchange-correlation interaction described by the generalized gradient approximation and parametrized by the Perdew-Burke-Ernzerhof functional with DFT-D$3$ van der Waals correction. Most of the available DFT simulations neglect the onset of nematicity, with notable exceptions which relate it to lattice distortions rather than to electronic correlations \cite{Subedi2022_PRM,Ptok2022_PRB}.

Taking even one further step back, before resolving the CO origin it is crucial to first determine the precise nature of the CO. Self-consistent mean-field theories such as Kohn-Sham DFT cannot unambiguously distinguish between a charge density wave (CDW), a state characterized by on-site order parameters, and a charge-bond order (CBO), which instead has intersite $\Delta_j$ components as its building blocks. Indeed, the presence of one of the two orders also induces the other, turning this into a chicken and egg problem.

A minimal model which is believed to encode the most salient properties of kagome metals is the single band extended Hubbard model defined on the kagome lattice at the (p-type) van Hove filling. Without interaction, this tight-binding model leads to the so-called sublattice interference mechanism. This means that parallel portions of the Fermi surface are characterized by a different sublattice index, which makes the local Hubbard interaction ineffective in inducing a finite wave vector ordered state which, in turn, leads to a more relevant role of the nearest-neighbor Coulomb repulsion \cite{Kiesel2012_PRB}. However, different methods applied to the study of this model provide very different results, with functional renormalization group suggesting the ground state to be a CBO driven by electronic instabilities \cite{Kiesel2013_PRL,Wang2013_PRB} while variational Monte Carlo indicates a critical role of phonons for the stabilization of this phase \cite{Ferrari2022_PRB}. Finally, self-consistent mean-field calculations point out the crucial role of the next-nearest neighbor Coulomb interaction, even if with unrealistically large magnitude, in getting a CO with broken TRS \cite{Zhou2022_NatComm,Dong2023_PRB}.

%%%%%%%%%%%%%%%%%%%%%%%%%%%%%%%

\begin{figure*}
\centerline{\includegraphics[width=1.0\textwidth]{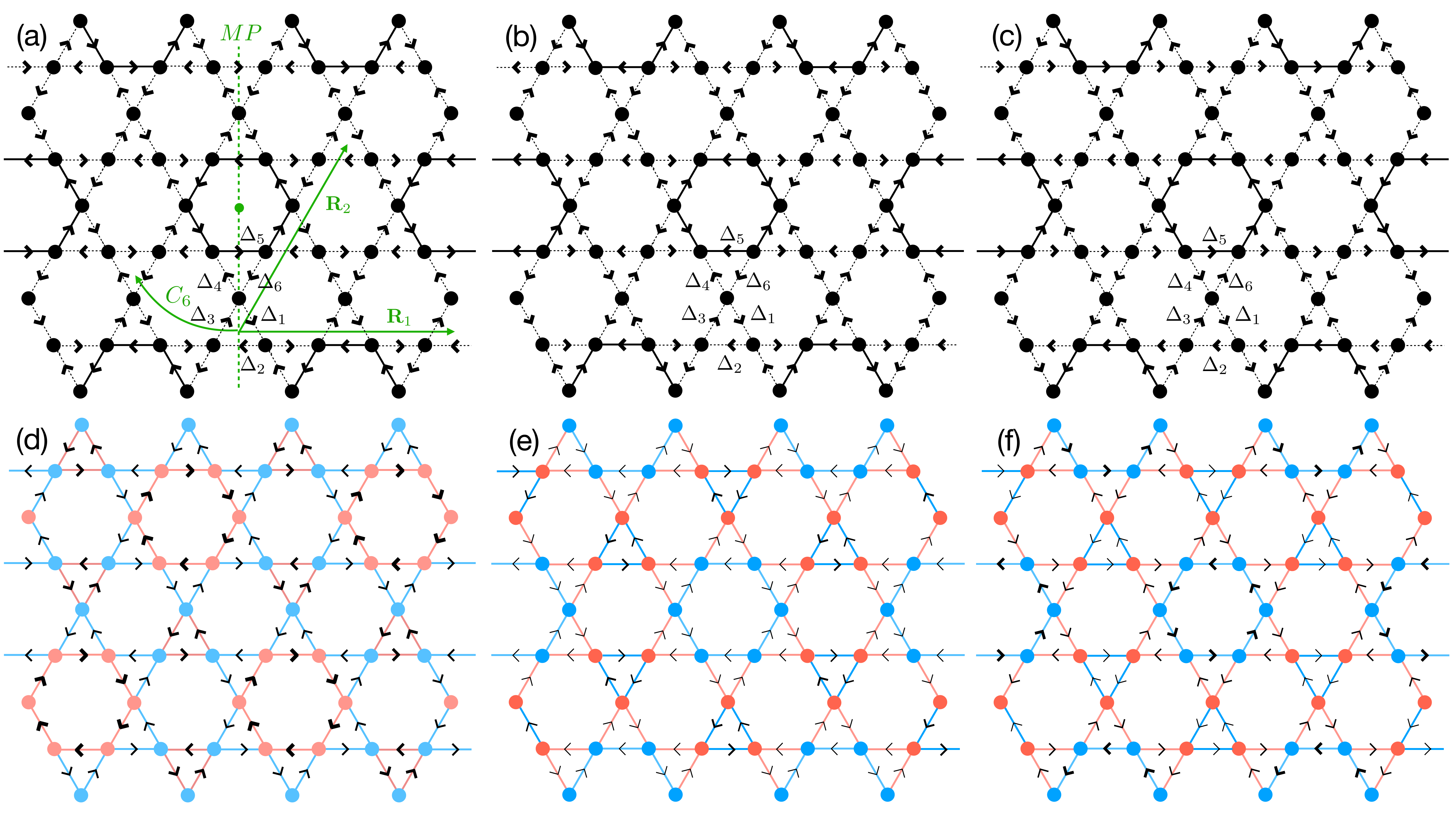}}
\caption{ \textbf{Hexagonal charge-bond order -} (a-c) Real-space representations of the Hex CBO, where solid (dashed) lines connecting two nearest neighbor atoms imply a positive (negative) sign for the real part of the order parameter $\mathcal{R}e \big[ \Delta_i \big]$. The arrows indicate the sign of the imaginary part $\mathcal{I}m \big[ \Delta_i \big]$, which is positive (negative) if we move along (opposite to) the direction of the arrow. (a) p$6$ wallpaper group pattern and the corresponding generating symmetry operations (in green) which consist of two translations $\mathbf{R}_1 : (\Delta_1, \Delta_2, \Delta_3, \Delta_4, \Delta_5, \Delta_6) \rightarrow (\Delta_1, \Delta_2, \Delta_3, \Delta_4, \Delta_5, \Delta_6)$ and $\mathbf{R}_2 : (\Delta_1, \Delta_2, \Delta_3, \Delta_4, \Delta_5, \Delta_6) \rightarrow (\Delta_1, \Delta_2, \Delta_3, \Delta_4, \Delta_5, \Delta_6)$ (which act as the identity), a rotation of $2 \pi / 6$ around the centre of the hexagonal pattern C$_6 : (\Delta_1, \Delta_2, \Delta_3, \Delta_4, \Delta_5, \Delta_6) \rightarrow (\Delta_3, \Delta_1, \Delta_2, \Delta_6, \Delta_4, \Delta_5)$, a mirror symmetry with respect to the dashed line shown in the image $\text{MP} : (\Delta_1, \Delta_2, \Delta_3, \Delta_4, \Delta_5, \Delta_6) \rightarrow (\Delta_3^*, \Delta_2^*, \Delta_1^*, \Delta_6^*, \Delta_5^*, \Delta_4^*)$ and $\text{TRS} : (\Delta_1, \Delta_2, \Delta_3, \Delta_4, \Delta_5, \Delta_6) \rightarrow (\Delta_1^*, \Delta_2^*, \Delta_3^*, \Delta_4^*, \Delta_5^*, \Delta_6^*)$. These transformations have to be supplied with the condition $(\Delta_4, \Delta_5, \Delta_6) \rightarrow (\Delta_1, - \Delta_2^*, \Delta_3)$, implying that the six parameters are not independent, but they can be reduced to three order parameters. (b) p$31$m wallpaper group pattern with symmetries C$_6 : (\Delta_1, \Delta_2, \Delta_3, \Delta_4, \Delta_5, \Delta_6) \rightarrow (\Delta_3^*, \Delta_1^*, \Delta_2^*, \Delta_6^*, \Delta_4^*, \Delta_5^*)$, $\text{MP} : (\Delta_1, \Delta_2, \Delta_3, \Delta_4, \Delta_5, \Delta_6) \rightarrow (\Delta_3^*, \Delta_2^*, \Delta_1^*, \Delta_6^*, \Delta_5^*, \Delta_4^*)$ and $\text{TRS} : (\Delta_1, \Delta_2, \Delta_3, \Delta_4, \Delta_5, \Delta_6) \rightarrow (\Delta_1^*, \Delta_2^*, \Delta_3^*, \Delta_4^*, \Delta_5^*, \Delta_6^*)$, together with $(\Delta_4, \Delta_5, \Delta_6) \rightarrow (\Delta_1, \Delta_2^*, \Delta_3)$. (c) p$6$ configuration with symmetry transformations C$_6 : (\Delta_1, \Delta_2, \Delta_3, \Delta_4, \Delta_5, \Delta_6) \rightarrow (\Delta_3, \Delta_1, \Delta_2, \Delta_6, \Delta_4, \Delta_5)$, $\text{MP} : (\Delta_1, \Delta_2, \Delta_3, \Delta_4, \Delta_5, \Delta_6) \rightarrow (\Delta_3^*, \Delta_2^*, \Delta_1^*, \Delta_6^*, \Delta_5^*, \Delta_4^*)$, $\text{TRS} : (\Delta_1, \Delta_2, \Delta_3, \Delta_4, \Delta_5, \Delta_6) \rightarrow (\Delta_1^*, \Delta_2^*, \Delta_3^*, \Delta_4^*, \Delta_5^*, \Delta_6^*)$ and $(\Delta_4, \Delta_5, \Delta_6) \rightarrow (\Delta_1^*, - \Delta_2^*, \Delta_3^*)$. (d-f) Real-space representation of the zero-temperature mean-field ($U=1.5$, $V=0.8$) occupation per site $\langle n_i \rangle$, the bond correlation pattern $\vert \langle c^\dagger_i c_{j \ \text{n.n.} \ i} \rangle \vert$ and of the current distribution corresponding to the order parameter configuration displayed in the panel above. Dark red corresponds to a large occupation (strong bond), while dark blue corresponds to a small occupation (weak bond). The size of the arrows suggests the magnitude of the current flowing through that bond. (d-f) $\psi_j = 0.3$, $\phi_1 = \phi_2 = \phi_3 = \pi/2$.}
\label{fig:hop2x2_Fe}
\end{figure*}

Given the above-explained controversy about the origin of charge order in kagome metals, and the fact that even a delicate interplay between electronic correlations and electron-phonon coupling might be at play, as is the case in other vanadium-based compounds such as VO$_2$ \cite{Grandi2020}, we take here a phenomenological perspective that disregards the origin of the ordered state starting from a minimal set of hypotheses concerning the CO. We assume:
\begin{itemize}
    \item[(1)] a $3$Q ordering, i.e., the presence of three order parameters $\Delta_1$, $\Delta_2$ and $\Delta_3$;
    \item[(2)] an in-plane $2 \times 2$ ($\sqrt{3} \times \sqrt{3}$) reconstruction;
    \item[(3)] unbroken point group symmetries of the lattice in the high-temperature metal (unless otherwise specified).
\end{itemize}
The resulting Ginzburg-Landau (GL) potentials host solutions with both nematic character and broken TRS \cite{Lin2021_PRB,Park2021_PRB,Denner2021_PRL,Yang2022_arXiv}. Particularly, the $3$Q ordering is crucial for the onset of nematicity since it permits to lower the C$_6$ symmetry of the system even if we assume $| \Delta_1 | = | \Delta_2 | = | \Delta_3 |$ at the instability level due to high order terms in the GL expansion. We stress that, in the framework provided by the single band Hubbard model, the order parameters of the CDW couple to the local electronic number operator on site $j$, $\Delta_j n_j$. Since $n_j$ is a Hermitian operator, in this case $\Delta_j$ would be real, impeding the breaking of TRS \cite{Nayak2000_PRB}. On the other hand, for the CBO the order parameters couple to the nearest-neighbor electronic hopping (which is not a Hermitian operator), allowing for a TRS broken state. For this reason, we believe that CBO is a more natural candidate state to describe the physics of kagome metals than the onsite CDW.

Furthermore, we analyze the real space form of the CBO induced by the three fields $\Delta_1$, $\Delta_2$ and $\Delta_3$ and by electronic correlations. Although a similar analysis has been already performed for kagome metals \cite{Denner2021_PRL}, we aim here to generalize it by taking into account all the $2 \times 2$ CBOs that have been suggested in the literature, such as the hexagonal (Hex), the tri-hexagonal (TrH) and the star of David (SoD) (together with their anti- partners), see the patterns in Figs.~\ref{fig:hop2x2_Fe}-\ref{fig:hop2x2_SoD}a \cite{Feng2021_PRB,Luo2022_npj,Liu2022_NatComm,Wang2022_arXiv,Han2023_AdvMat,Uykur2021_PRB,Uykur2022_npj,Wang2022_PRB,Tsirlin2022_SciPost,Luo2022_NatComm,Tan2021_PRL,Ortiz2021_PRX,Miao2021_PRB,Consiglio2022_PRB,Dong2023_PRB}. Besides that, we consider also the $\sqrt{3} \times \sqrt{3}$ CBO, see the pattern in Fig.~\ref{fig:hopsq3xsq3}a \cite{Chen2021_Nat,Arachchige2022_PRL,Hu2022_arXiv,Ferrari2022_PRB}.

The article is structured as follows. In Sec.~\ref{sec:GL+MF}, we study several CBO patterns and their corresponding GL potentials. We analyze in which regime of parameters the GL 
%FG
free energy 
%free-energy 
might support the onset of TRS breaking and of nematicity, and we perform a mean-field calculation providing a real-space representation of the corresponding CBO. We find that nematicity can occur in two ways: through a different phase, and through a different amplitude of the complex order parameters $\Delta_1$, $\Delta_2$ and $\Delta_3$. In Sec.~\ref{sec:RUS}, we discuss how resonant ultrasound spectroscopy (RUS) can be applied in the present context and which information one might extract from it  \cite{Ghosh2020_SciAdv,Ghosh2021_Nat_Phys,Benhabib2021_NatPhys}. Particularly, we show that RUS can distinguish if the order parameters have one or two components and that it might provide information regarding the critical temperature $T_\text{nem}$ for the onset of the nematic CBO. Finally, Sec.~\ref{sec:Conclusions} is devoted to discussing theoretical and experimental implications of our results and to concluding remarks.

%%%%%%%%%%%%%%%%%%%%%%%%%%%%%%%%%%%%%%%%%%%%%%% GL+MF
\section{Ginzburg-Landau and mean-field analysis} \label{sec:GL+MF}
\begin{figure*}
\centerline{\includegraphics[width=1.0\textwidth]{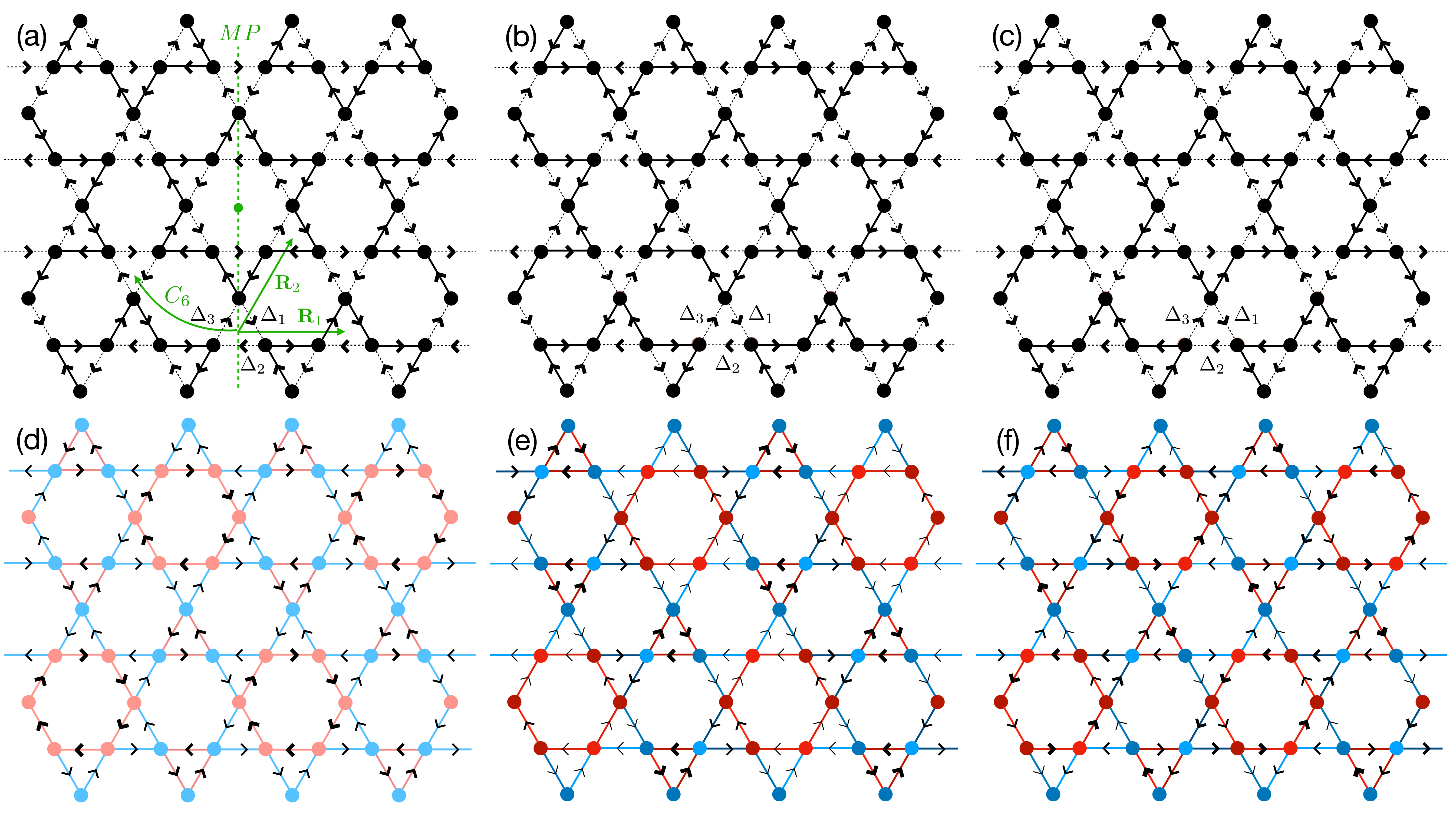}}
\caption{ \textbf{Tri-hexagonal charge-bond order -} (a-c) Real-space representations of the TrH CBO, where the same notation used in Figs.~\ref{fig:hop2x2_Fe}a-c is used. (a) p$6$ wallpaper group pattern \cite{Dong2023_PRB} and the corresponding generating symmetry operations (in green) which consist of two translations $\mathbf{R}_1 : (\Delta_1, \Delta_2, \Delta_3) \rightarrow (- \Delta_1^*, \Delta_2, - \Delta_3^*)$ and $\mathbf{R}_2 : (\Delta_1, \Delta_2, \Delta_3) \rightarrow (- \Delta_1^*, - \Delta_2^*, \Delta_3)$, a rotation of $2 \pi / 6$ around the axis orthogonal to the plane of the lattice passing through the centre of the anti-tri-hexagonal pattern C$_6 : (\Delta_1, \Delta_2, \Delta_3) \rightarrow (\Delta_3, \Delta_1, \Delta_2)$, a mirror symmetry with respect to a plane orthogonal to the plane of the lattice passing through the green dashed line shown in the image $\text{MP} : (\Delta_1, \Delta_2, \Delta_3) \rightarrow (\Delta_3^*, \Delta_2^*, \Delta_1^*)$ and the time reversal symmetry $\text{TRS} : (\Delta_1, \Delta_2, \Delta_3) \rightarrow (\Delta_1^*, \Delta_2^*, \Delta_3^*)$. (b) p$31$m wallpaper group pattern \cite{Denner2021_PRL} with symmetries $\mathbf{R}_1 : (\Delta_1, \Delta_2, \Delta_3) \rightarrow (- \Delta_1, \Delta_2, - \Delta_3)$, $\mathbf{R}_2 : (\Delta_1, \Delta_2, \Delta_3) \rightarrow (- \Delta_1, - \Delta_2, \Delta_3)$, C$_6 : (\Delta_1, \Delta_2, \Delta_3) \rightarrow (\Delta_3^*, \Delta_1^*, \Delta_2^*)$, $\text{MP} : (\Delta_1, \Delta_2, \Delta_3) \rightarrow (\Delta_3^*, \Delta_2^*, \Delta_1^*)$ and $\text{TRS} : (\Delta_1, \Delta_2, \Delta_3) \rightarrow (\Delta_1^*, \Delta_2^*, \Delta_3^*)$. (c) p$6$ configuration \cite{Feng2021_sci_brief,Park2021_PRB,Yang2022_arXiv,Dong2023_PRB} with symmetry transformations $\mathbf{R}_1 : (\Delta_1, \Delta_2, \Delta_3) \rightarrow (- \Delta_1, \Delta_2, - \Delta_3)$, $\mathbf{R}_2 : (\Delta_1, \Delta_2, \Delta_3) \rightarrow (- \Delta_1, - \Delta_2, \Delta_3)$, C$_6 : (\Delta_1, \Delta_2, \Delta_3) \rightarrow (\Delta_3, \Delta_1, \Delta_2)$, $\text{MP} : (\Delta_1, \Delta_2, \Delta_3) \rightarrow (\Delta_3^*, \Delta_2^*, \Delta_1^*)$ and $\text{TRS} : (\Delta_1, \Delta_2, \Delta_3) \rightarrow (\Delta_1^*, \Delta_2^*, \Delta_3^*)$. (d-f) Same quantities and values as described in the caption of Fig.~\ref{fig:hop2x2_Fe}. (d) $\psi_j = 0.3$, $\phi_1 = \phi_2 = \phi_3 = \pi/2$, (e-f) $\psi_j = 0.3$, $\phi_1 = 2.4$, $\phi_2 = \phi_3 = 0.7$.}
\label{fig:hop2x2}
\end{figure*}

%FG
The precise form of the $2 \times 2$ in-plane modulation of the kagome lattice in the kagome metals $A$V$_3$Sb$_5$ ($A=$K, Rb, Cs) is not known. For this reason, we describe all the proposals that have been, to the best of our knowledge, advanced for this phase.
%The precise form of the $2 \times 2$ in-plane modulation of the kagome lattice in the kagome metals $A$V$_3$Sb$_5$ ($A=$K, Rb, Cs) is not known. For this reason, we take an unbiased viewpoint, trying to describe, within a phenomenological theory, all the proposals that have been, to the best of our knowledge, advanced for the specific form of this phase. 
We consider a modulation of real-space hoppings that preserves the mirror C$_6$ symmetries of the lattice. In this respect, there are three possibilities: a Hex (Fig.~\ref{fig:hop2x2_Fe}a), a TrH (Fig.~\ref{fig:hop2x2}a) and a SoD (Fig.~\ref{fig:hop2x2_SoD}a) patterns and their corresponding ``anti-'' partners (which, however, do not differ from the point of view of the GL potential) \cite{Feng2021_PRB}. A similar possibility is analyzed for a different ordering vector, which leads to a $\sqrt{3} \times \sqrt{3}$ unit cell (Fig.~\ref{fig:hopsq3xsq3}a). 
%FG
Even in this case, no assumptions on the origin of the ordered state are made. 
Since the kagome lattice has three independent directions, each rotated by $\pm 2 \pi/3$ with respect to the other, we can define three independent order parameters $\Delta_1$, $\Delta_2$ and $\Delta_3$, represented, for instance, in Figs.~\ref{fig:hop2x2}a and \ref{fig:hopsq3xsq3}a  \cite{Kiesel2013_PRL}. In Figs.~\ref{fig:hop2x2_Fe}a-c and Figs.~\ref{fig:hop2x2_SoD}a-c, six $\Delta$ parameters are displayed, however $\Delta_4$, $\Delta_5$ and $\Delta_6$ are dependent by $\Delta_1$, $\Delta_2$ and $\Delta_3$ (see the corresponding captions for further details).

Since the CBO might break the TRS, we allow the order parameters to be complex. The sign of the phase acquired by an electron during a hopping process is given by the direction of the arrows shown in Figs.~\ref{fig:hop2x2_Fe}-\ref{fig:hopsq3xsq3}a-c. For the $2 \times 2$ unit cell, we consider three configurations, already investigated in the literature \cite{Denner2021_PRL,Feng2021_sci_brief,Park2021_PRB,Yang2022_arXiv,Dong2023_PRB}, for the imaginary part of the order parameters, each of them represented in panels a, b and c, respectively, of Figs.~\ref{fig:hop2x2_Fe}-\ref{fig:hop2x2_SoD}. Despite the fact that for the $\sqrt{3} \times \sqrt{3}$ CO there is currently no evidence for the breaking of the TRS, we allow this possibility by suggesting two patterns for the imaginary hoppings represented in Figs.~\ref{fig:hopsq3xsq3}b-c.

The general expression for the GL free energy depends on the three order parameters as $\mathcal{F} \propto h \sum_j \Delta_j + \alpha \sum_{j,k} \Delta_j \Delta_k + \gamma \sum_{j,k,l} \Delta_j \Delta_k \Delta_l + \beta \sum_{j,k,l,m} \Delta_j \Delta_k \Delta_l \Delta_m$, having arrested the expansion to the fourth order, neglected the gradient terms and disregarded the complex conjugations for brevity. Panel a of Figs.~\ref{fig:hop2x2_Fe}-\ref{fig:hop2x2_SoD} and panel b of Fig.~\ref{fig:hopsq3xsq3} show the generating symmetries of the corresponding pattern, consisting in two lattice translations $\mathbf{R}_1$ and $\mathbf{R}_2$, a sixfold rotation C$_6$ and a mirror plane $\text{MP}$. In addition to them, we also consider the $\text{TRS}$. 
%FG
%The Hex and the SoD patterns are also characterized by a transformation of $\Delta_4$, $\Delta_5$ and $\Delta_6$ into $\Delta_1$, $\Delta_2$ and $\Delta_3$. 
These symmetry operations might force some of the contributions to the potential to be zero, e.g.\ the linear or the cubic terms. Next, we analyze separately each of the above mentioned patterns for the CBO.

%%%%%%% 2x2 UNIT CELL: Hexagonal
\subsection{ $2 \times 2$ unit cell: Hexagonal CBO}
%FG
We start by analyzing the Hex CBO (Figs.~\ref{fig:hop2x2_Fe}a-c). 
%We start by analyzing the Hex CBO depicted in Figs.~\ref{fig:hop2x2_Fe}a-c. 
By writing each of the order parameters in terms of amplitude and phase $\Delta_j = \psi_j e^{i \phi_j}$, $j=1,2,3$ and $\psi_j > 0$, we arrive at the expression of the GL potential \cite{vanWezel2011_EPL}:
\begin{align} \label{GL_0}
	\mathcal{F}_{\text{Hex}} & = \sum_j \psi_j^2 \big[ \alpha_1 + \alpha_2 \cos (2 \phi_j) + \frac{\beta}{4} \psi_j^2 \big] ,
\end{align}
where the quartic interaction has been included in its simplest time-reversal invariant form (the general expression of this interaction is provided in the Supplemental 
%FG
Material \cite{Suppl_mat}).
%Material). 
The combination of the contributions proportional to $\alpha_1$ and $\beta$ is characteristic of most of the GL potentials; the stabilization of a state with finite (zero) amplitude $\psi_j$ is assured by considering $\beta>0$ and $\alpha_1 < 0$ ($\alpha_1 > 0$) below (above) the critical temperature $T_\text{C}$ of the system; $\alpha_1 \propto T - T_\text{C}$. The term proportional to $\alpha_2$ is minimized by $\phi_j \ \text{mod} \ \pi = 0$ for $\alpha_2 <0$ and by $\phi_j \ \text{mod} \ \pi = \pi/2$ for $\alpha_2 > 0$. Thus, the potential in Eq.~\eqref{GL_0} has only two non-trivial solutions: one in which the order parameters are purely real and one in which they are purely imaginary. One can find an analytic expression for the minima of Eq.~\eqref{GL_0}, with $\psi_j = \sqrt{\frac{2 ( | \alpha_2 | - \alpha_1)}{\beta}}$, implying $| \alpha_2 | > \alpha_1$ below the critical temperature. Since our primary goal is to stabilize a (nematic) state that breaks the TRS, we assume $\alpha_2 > 0$; a condition that leads to purely imaginary order parameters. With these assumptions, we do not have to distinguish between the Hex (as represented in Figs.~\ref{fig:hop2x2_Fe}a-c) and the anti-Hex configurations.

To obtain additional information about the symmetry properties of the resulting state we perform a mean-field analysis of the extended Hubbard model on the kagome lattice in the presence of the ordering fields $\Delta_j$, $j=1,2,3$ \cite{Jiang2021_Nat_Mat,Zhou2022_NatComm}. The Hamiltonian of the problem is:
\begin{align} \label{ham}
	H' & = -t \sum_{\langle i,j \rangle, \sigma} \big( c^\dagger_{i,\sigma} c_{j, \sigma} + \text{H.c.} \big) + \mu \sum_i n_i \nonumber \\
	& + U \sum_i n_{i, \uparrow} n_{i, \downarrow} + V \sum_{\langle i,j \rangle} n_i n_j , 
\end{align}
where $c^\dagger_{i,\sigma}$ ($c_{i,\sigma}$) corresponds to the creation (annihilation) operator for one electron on site $i$ and with spin $\sigma$ and $n_i = n_{i, \uparrow} + n_{i, \downarrow}$ is the total occupation for site $i$ (given by the sum of the occupations per spin). In Eq.~\eqref{ham}, $t$ is the nearest-neighbor hopping integral ($t=1$ sets our energy scale), $\mu$ is the chemical potential, which is fixed to have $2.5$ electrons every three sites (filling fraction $5/6$, corresponding to the p-type filling \cite{Kiesel2012_PRB}), $U$ sets the magnitude of the local Hubbard repulsion, and $V$ is the nearest-neighbor Coulomb interaction. We consider the Hamiltonian \eqref{ham} to be renormalized by the $\Delta_j$, so
\begin{align} \label{ham_renorm}
    H = H' + \sum_{\mathbf{R}} \boldsymbol{\Delta}(\mathbf{R}) \cdot \hat{\mathbf{O}} (\mathbf{R}) ,
\end{align}
with $\hat{\mathbf{O}} (\mathbf{R})$ the vector that contains all the $24$ ($18$) operators of the form $\sum_\sigma c^\dagger_{i,\sigma} c_{j,\sigma}$ in the $2 \times 2$ ($\sqrt{3} \times \sqrt{3}$) unit cell at $\mathbf{R}$, with $i$ and $j$ nearest-neighbor sites, and $\boldsymbol{\Delta}(\mathbf{R})$ the vector containing the corresponding values of $\Delta_1$, $\Delta_2$ and $\Delta_3$ as represented in Figs.~\ref{fig:hop2x2_Fe}a-c.

The mean-field solutions of the Hamiltonian \eqref{ham_renorm} are shown in Figs.~\ref{fig:hop2x2_Fe}d-f for the patterns represented in Figs.~\ref{fig:hop2x2_Fe}a-c, respectively (additional details on the mean-field procedure are provided in the Supplemental Material \cite{Suppl_mat,Wen2010_PRB,Liu2010_PRB,Lopez2020_PRB}). In all these cases, the CDW and the CBO do not break the rotational and the mirror symmetry of the problem, and they can be identified as TrH (Fig.~\ref{fig:hop2x2_Fe}d) and anti-TrH (Figs.~\ref{fig:hop2x2_Fe}e and f) configurations. Moreover, the imaginary order parameters produce finite currents in the lattice which might reduce the rotational symmetry of the problem, as it is the case for Figs.~\ref{fig:hop2x2_Fe}e-f, that are C$_3$-symmetric. None of the configurations reached is nematic.

%%%%%%% 2x2 UNIT CELL: Tri-hexagonal
\subsection{ $2 \times 2$ unit cell: Tri-hexagonal CBO}
The 
%FG
configuration we analyze next is of the TrH (or anti-TrH) type \cite{Feng2021_PRB} (Fig.~\ref{fig:hop2x2}a), leading 
%configurations we analyze next are of the TrH (or anti-TrH) type \cite{Feng2021_PRB}, with generating symmetries shown in Figs.~\ref{fig:hop2x2}a. They lead 
to the GL potential:
\begin{align} \label{GL_1}
	\mathcal{F}_\text{TrH}^\text{a} & = \sum_j \psi_j^2 \big[ \alpha_1 + \alpha_2 \cos (2 \phi_j) + \frac{\beta}{4} \psi_j^2 \big]  \nonumber \\
	& + \frac{8 \gamma}{3} \psi_1 \psi_2 \psi_3 \cos (\phi_1) \cos (\phi_2) \cos (\phi_3) .
\end{align}
When $\gamma = 0$, Eq.~\eqref{GL_1} reduces to Eq.~\eqref{GL_0}. The minimization of the cubic interaction leads to $\psi_1 = \psi_2 = \psi_3$ and $(\phi_1,\phi_2,\phi_3) \ \text{mod} \ (2 \pi, 2 \pi, 2 \pi) = (\pi,\pi,\pi), (\pi,0,0), (0,\pi,0)$ or $(0,0,\pi)$ when $\gamma > 0$, while, for $\gamma < 0$, the minima are  $(\phi_1,\phi_2,\phi_3) \ \text{mod} \ (2 \pi, 2 \pi, 2 \pi) = (0,0,0), (\pi,\pi,0), (\pi,0,\pi)$ or $(0,\pi,\pi)$. If we assume $\alpha_2 > 0$ and $\alpha_2 \gg |\gamma|$, the analytic expression found for Eq.~\eqref{GL_0} also holds for the potential Eq.~\eqref{GL_1} (indeed, in this case, we assume an irrelevant contribution coming from the cubic term, effectively mapping Eq.~\eqref{GL_1} into Eq.~\eqref{GL_0}). The mean-field solution of the Hamiltonian in Eq.~\eqref{ham_renorm} with spatial distribution of the order parameters shown in Fig.~\ref{fig:hop2x2}a, is represented in Fig.~\ref{fig:hop2x2}d, and it is found to be identical to Fig.~\ref{fig:hop2x2_Fe}d, consistent with the purely imaginary nature of the order parameters in the two cases. As a side remark, we notice that the potential in Eq.~\eqref{GL_1} can describe a CDW with $2 \times 2$ unit cell provided that we assume $\Delta_j$ to be real valued, i.e. $\phi_j \ \text{mod} \ \pi = 0$, and that they are local order parameters.

\begin{figure*}
\centerline{\includegraphics[width=1.0\textwidth]{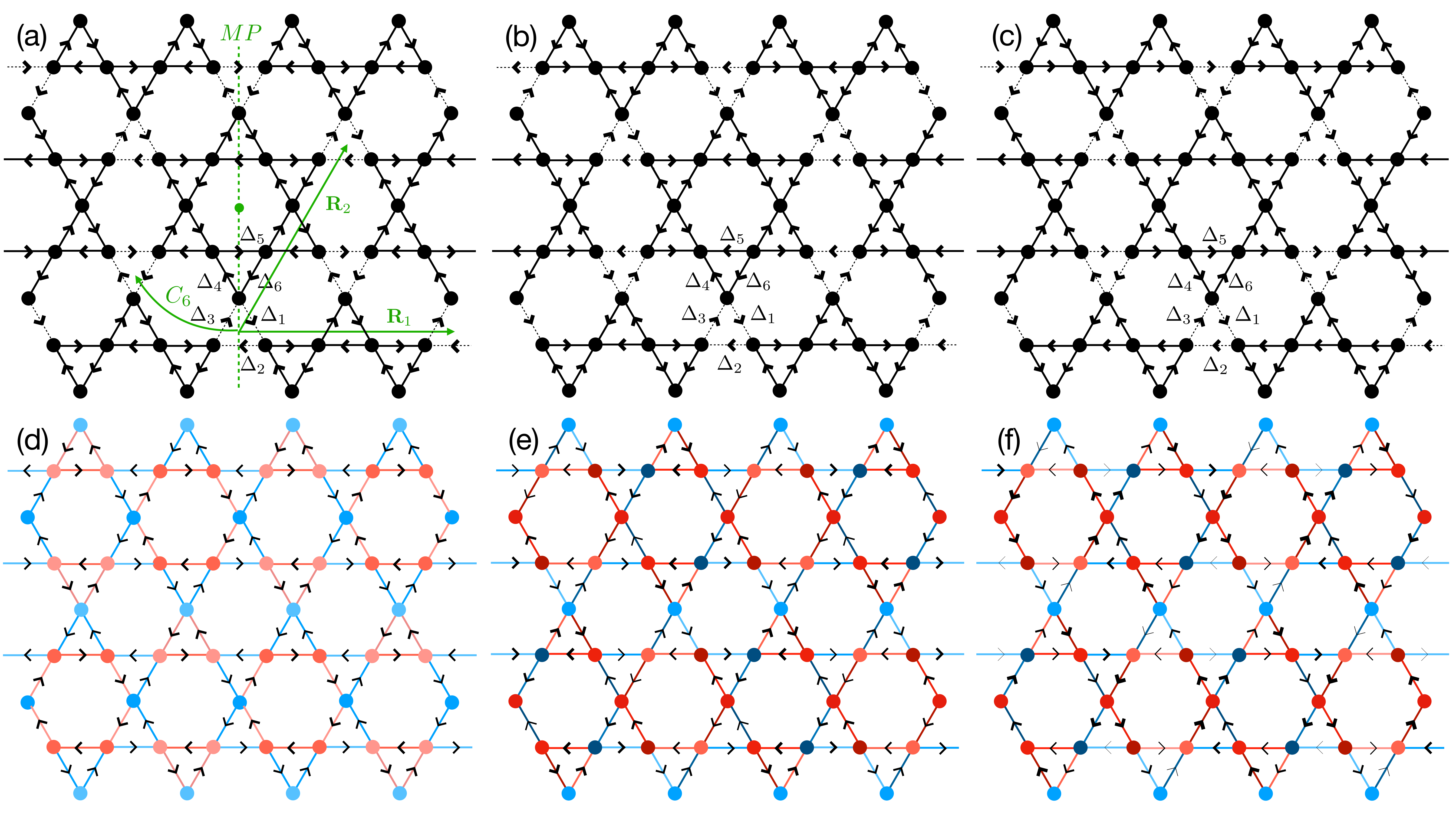}}
\caption{ \textbf{Star of David charge-bond order -} (a-c) Real-space representations of the SoD CBO, where the same convention introduced in Figs.~\ref{fig:hop2x2_Fe}a-c is used. The translations act in a trivial way ($\mathbf{R}_1 : (\Delta_1, \Delta_2, \Delta_3, \Delta_4, \Delta_5, \Delta_6) \rightarrow (\Delta_1, \Delta_2, \Delta_3, \Delta_4, \Delta_5, \Delta_6)$ and $\mathbf{R}_2 : (\Delta_1, \Delta_2, \Delta_3, \Delta_4, \Delta_5, \Delta_6) \rightarrow (\Delta_1, \Delta_2, \Delta_3, \Delta_4, \Delta_5, \Delta_6)$). (a) p$6$ wallpaper group pattern with generating symmetries C$_6 : (\Delta_1, \Delta_2, \Delta_3, \Delta_4, \Delta_5, \Delta_6) \rightarrow (\Delta_3, \Delta_1, \Delta_2, \Delta_6, \Delta_4, \Delta_5)$, $\text{MP} : (\Delta_1, \Delta_2, \Delta_3, \Delta_4, \Delta_5, \Delta_6) \rightarrow (\Delta_3^*, \Delta_2^*, \Delta_1^*, \Delta_6^*, \Delta_5^*, \Delta_4^*)$, $\text{TRS} : (\Delta_1, \Delta_2, \Delta_3, \Delta_4, \Delta_5, \Delta_6) \rightarrow (\Delta_1^*, \Delta_2^*, \Delta_3^*, \Delta_4^*, \Delta_5^*, \Delta_6^*)$ and the mapping $(\Delta_4, \Delta_5, \Delta_6) \rightarrow (-\Delta_1^*, - \Delta_2^*, -\Delta_3^*)$. (b) p$31$m wallpaper group pattern with symmetries C$_6 : (\Delta_1, \Delta_2, \Delta_3, \Delta_4, \Delta_5, \Delta_6) \rightarrow (\Delta_3^*, \Delta_1^*, \Delta_2^*, \Delta_6^*, \Delta_4^*, \Delta_5^*)$, $\text{MP} : (\Delta_1, \Delta_2, \Delta_3, \Delta_4, \Delta_5, \Delta_6) \rightarrow (\Delta_3^*, \Delta_2^*, \Delta_1^*, \Delta_6^*, \Delta_5^*, \Delta_4^*)$ and $\text{TRS} : (\Delta_1, \Delta_2, \Delta_3, \Delta_4, \Delta_5, \Delta_6) \rightarrow (\Delta_1^*, \Delta_2^*, \Delta_3^*, \Delta_4^*, \Delta_5^*, \Delta_6^*)$, together with $(\Delta_4, \Delta_5, \Delta_6) \rightarrow (-\Delta_1^*, -\Delta_2, -\Delta_3^*)$. (c) p$6$ configuration with symmetry transformations C$_6 : (\Delta_1, \Delta_2, \Delta_3, \Delta_4, \Delta_5, \Delta_6) \rightarrow (\Delta_3, \Delta_1, \Delta_2, \Delta_6, \Delta_4, \Delta_5)$, $\text{MP} : (\Delta_1, \Delta_2, \Delta_3, \Delta_4, \Delta_5, \Delta_6) \rightarrow (\Delta_3^*, \Delta_2^*, \Delta_1^*, \Delta_6^*, \Delta_5^*, \Delta_4^*)$, $\text{TRS} : (\Delta_1, \Delta_2, \Delta_3, \Delta_4, \Delta_5, \Delta_6) \rightarrow (\Delta_1^*, \Delta_2^*, \Delta_3^*, \Delta_4^*, \Delta_5^*, \Delta_6^*)$ and $(\Delta_4, \Delta_5, \Delta_6) \rightarrow (-\Delta_1, - \Delta_2^*, -\Delta_3)$. (d-f) Same quantities and values as described in the caption of Fig.~\ref{fig:hop2x2_Fe}. (d) $\psi_1 = \psi_3 = 0$, $\psi_2 = 0.3$, $\phi_2 = \pi/2$. (e-f) $\psi_1 = \psi_3 = 0.35$, $\psi_2 = 0.20$, $\phi_1 = 0$, $\phi_2 = \pi/2$ and $\phi_3 = \pi$.}
\label{fig:hop2x2_SoD}
\end{figure*}

A different form of the GL potential can be obtained by analyzing the patterns in Figs.~\ref{fig:hop2x2}b and c. Both of them are characterized by the same symmetry constraints, going across two sign changes and three complex conjugations under the action of the generating symmetry operations, leading to the potential \cite{Lin2021_PRB,Park2021_PRB,Denner2021_PRL}:
\begin{align} \label{GL_2}
	& \mathcal{F}_\text{TrH}^\text{b} = \sum_j \psi_j^2 \big[ \alpha_1 + \alpha_2 \cos (2 \phi_j) + \frac{\beta}{4} \psi_j^2 \big] \\
    & + \frac{2}{3} \psi_1 \psi_2 \psi_3 \Big[ (\gamma_1 - \gamma_2) \cos \big( \sum_j \phi_j \big) + 4 \gamma_2 \prod_j \cos (\phi_j) \Big] \nonumber .
\end{align}
In passing, we notice that Eq.~\eqref{GL_2} reduces to Eq.~\eqref{GL_1} when $\gamma_1 = \gamma_2 = \gamma$. The term proportional to $\gamma_1$, when $\gamma_1 > 0$ ($\gamma_1 < 0$), is minimized by $( \phi_1 + \phi_2 + \phi_3 ) \ \text{mod} \ 2 \pi = \pi$ ($( \phi_1 + \phi_2 + \phi_3 ) \ \text{mod} \ 2 \pi = 0$), a condition that is satisfied, e.g., by $\phi_2 \neq \phi_{1,3} \neq \pi/2$, implying complex order parameters and a $\mathbb{Z}_3$ symmetry which has been suggested to be relevant for kagome metals \cite{Nie2022_Nat,Tazai2022_arXiv,Xu2022_NatPhys} and for twisted bilayer graphene \cite{Fernandes2020_SciAdv}. In the case where all $\phi_j \ \text{mod} \ \pi \neq 0$, the phase reached has also been called $3$\textbf{Q}$-3$\textbf{Q} (the three order parameters are all complex, having both real and imaginary components), while, in the case $\phi_2 \ \text{mod} \ \pi = 0$ and $\phi_{1,3} \ \text{mod} \ \pi = \pi/2$, the phase has been called $2$\textbf{Q}$-1$\textbf{Q} (two of the three order parameters are purely imaginary and one is purely real) \cite{Christensen2022_PRB}. These phases can thus be regarded as limiting cases of the condition $( \phi_1 + \phi_2 + \phi_3 ) \ \text{mod} \ 2 \pi = \pi$ enforced by $\gamma_1 > 0$. The term proportional to $\gamma_2$ in Eq.~\eqref{GL_2} tends to stabilize real order parameters, with $\phi_1 = \phi_2 = \phi_3 = \pi \ \text{mod} \ 2 \pi$ when $\gamma_2 > 0$ and $\phi_1 = \phi_2 = \phi_3 = 0 \ \text{mod} \ 2 \pi$ when $\gamma_2 < 0$. Both the cubic terms concur in realizing a state with $\psi_1 = \psi_2 = \psi_3$. The presence of cubic interactions in Eqs.~\eqref{GL_1} and \eqref{GL_2} might justify the weak first-order character of the transition to the CO phase observed for the kagome metals $A$V$_3$Sb$_5$ ($A=$K, Rb, Cs).

For $\gamma_2 = 0$ \cite{Park2021_PRB,Denner2021_PRL,Yang2022_arXiv}, the potential in Eq.~\eqref{GL_2} describes the onset of a C$_2$-symmetric CBO with TrH shape, as confirmed by the mean-field results displayed in Figs.~\ref{fig:hop2x2}e and f, as well as by previous analysis \cite{Denner2021_PRL,Lin2021_PRB}. Despite the similarities of the CBO and the CDW in the two configurations (Figs.~\ref{fig:hop2x2}e and f), the distribution of the currents is very different. Particularly, while the currents displayed in Fig.~\ref{fig:hop2x2}e preserve a mirror symmetry plane but break inversion symmetry (C$_\text{S}$-symmetric), the ones shown in Fig.~\ref{fig:hop2x2}f preserve inversion symmetry but do not have any mirror plane (C$_2$-symmetric). In this respect, only the state in Fig.~\ref{fig:hop2x2}f is nematic.

\begin{figure*}
\centerline{\includegraphics[width=1.0\textwidth]{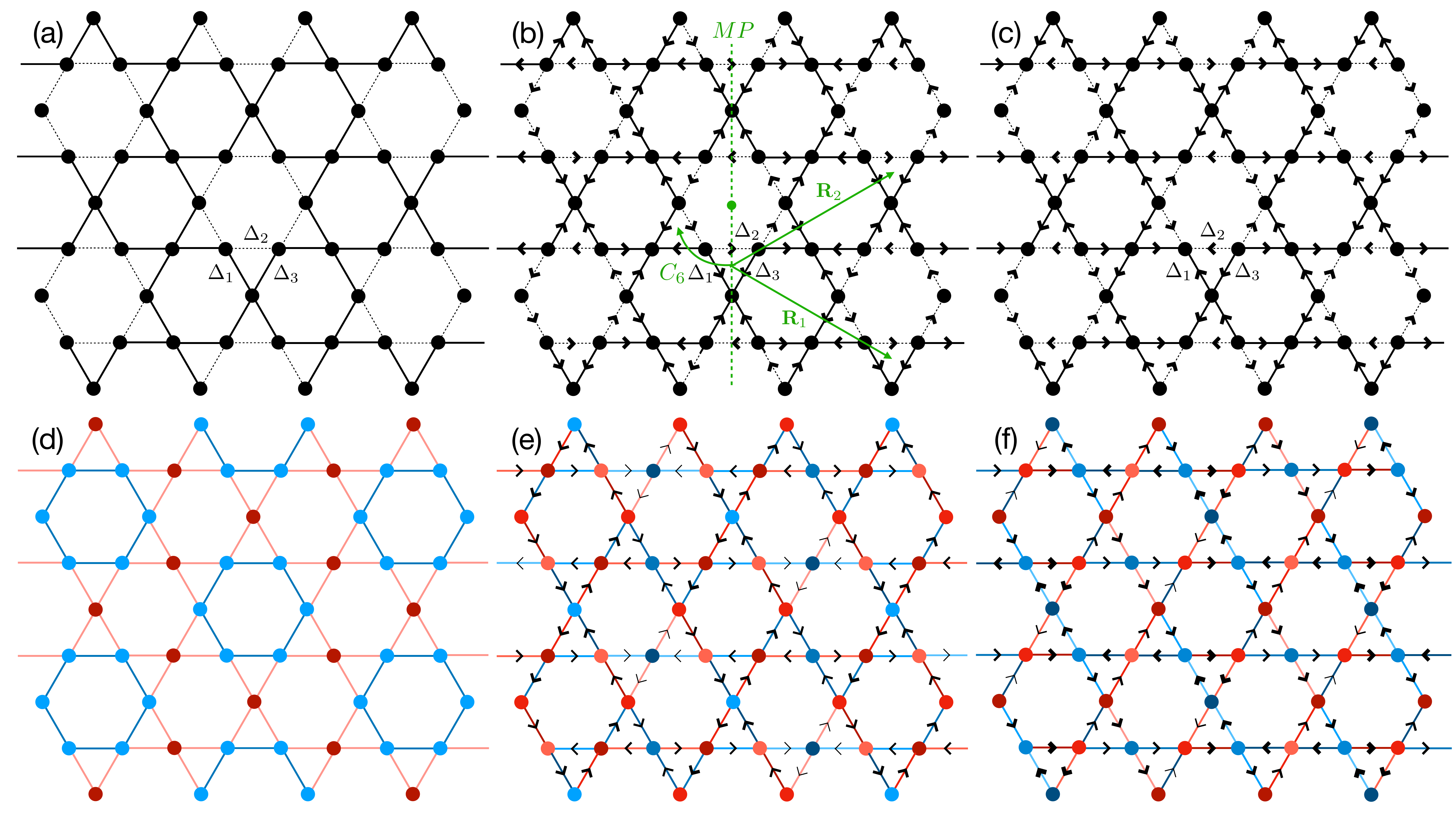}}
\caption{ \textbf{$\sqrt{3} \times \sqrt{3}$ charge-bond order -} (a-c) Real-space representations of the CBO with $\sqrt{3} \times \sqrt{3}$ unit cell. (a) Pattern belonging to the p$6$ wallpaper group with real order parameters $(\Delta_1, \Delta_2, \Delta_3)$. (b) The pattern belongs to the p$6$ wallpaper group, and the lattice translations $\mathbf{R}_1$ and $\mathbf{R}_2$ act in a trivial way on the order parameter. The sixfold rotation transforms the vector of the order parameters as C$_6 : (\Delta_1, \Delta_2, \Delta_3) \rightarrow (\Delta_3, \Delta_1, \Delta_2)$. The mirror plane symmetry acts as $\text{MP} : (\Delta_1, \Delta_2, \Delta_3) \rightarrow (\Delta_3, \Delta_2^*, \Delta_1)$ while the time reversal symmetry is $\text{TRS} : (\Delta_1, \Delta_2, \Delta_3) \rightarrow (\Delta_1^*, \Delta_2^*, \Delta_3^*)$. (c) Pattern belonging to the p$3$m$1$ wallpaper group, characterized by the transformations C$_6 : (\Delta_1, \Delta_2, \Delta_3) \rightarrow (\Delta_3^*, \Delta_1^*, \Delta_2^*)$, $\text{MP} : (\Delta_1, \Delta_2, \Delta_3) \rightarrow (\Delta_3^*, \Delta_2^*, \Delta_1^*)$ and $\text{TRS} : (\Delta_1, \Delta_2, \Delta_3) \rightarrow (\Delta_1^*, \Delta_2^*, \Delta_3^*)$. (d-f) Same quantities and values as described in the caption of Fig.~\ref{fig:hop2x2_Fe}. (d) $\psi_1 = \psi_2 = \psi_3 = 0.3$, $\phi_1 = \phi_3 = \pi$, $\phi_2 = 0$, (e) $\psi_1 = \psi_3 = 0.26$, $\psi_2 = 0.34$, $\phi_1 = \phi_3 = 0$, $\phi_2 = 1.72$, (f) $\psi_1 = \psi_3 = 0.26$, $\psi_2 = 0.34$, $\phi_1 = \phi_3 = -1.496$, $\phi_2 = 2.43$. Panels (a) and (d), which display a real order parameter configuration, are shown as a reference.}
\label{fig:hopsq3xsq3}
\end{figure*}

%%%%%%% 2x2 UNIT CELL: Star of David
\subsection{ $2 \times 2$ unit cell: Star of David CBO}
We now move to the discussion of the $2 \times 2$ CBO with SoD shape, as displayed in Figs.~\ref{fig:hop2x2_SoD}a-c. The spatial distribution of the order parameters depicted in Fig.~\ref{fig:hop2x2_SoD}a leads to the GL potential:
\begin{align} \label{GL_3}
	& \mathcal{F}_\text{SoD}^\text{a} = \sum_j \psi_j^2 \big[ \alpha_1 + \alpha_2 \cos (2 \phi_j) + \frac{\beta}{4} \psi_j^2 \big] \nonumber \\
	& + \sum_{j, j' > j} \psi_j \psi_{j'} \big[ \alpha_3 \cos (\phi_j + \phi_{j'}) + \alpha_4 \cos (\phi_j - \phi_{j'}) \big] .
\end{align}
The contribution to the potential proportional to $\alpha_3$ favors the solution $\phi_j \ \text{mod} \ \pi = 0$, $\psi_1 = \psi_2 = \psi_3$ when $\alpha_3 < 0$; if instead $\alpha_3 > 0$, we generally get $\phi_1 = \phi_2 = \phi_3 = \pi /2 \ \text{mod} \ \pi$ and $\psi_1 = \psi_2 = \psi_3$ at the minimum (something similar is found for the contribution proportional to $\alpha_4$). Thus, in the case $\alpha_4 = 0$, $\alpha_1, \alpha_3 < 0$, $\alpha_2, \beta > 0$, the competition among the quadratic contributions $\alpha_2$ and $\alpha_3$ might stabilize a solution with $\psi_1 = \psi_3 < \psi_2$ and $\phi_2 \ \text{mod} \ 2 \pi = - \pi/2$, $\phi_1 = \phi_3 = \pi/2 \ \text{mod} \ 2 \pi$. Taking this behaviour to its extremes, we might assume $\psi_1 = \psi_3 = 0$, while $\psi_2$ is still finite (this assumption is consistent with the presence of an interaction $\beta (\psi_1^2 + \psi_2^2 + \psi_3^2)^2 /4$ instead of the quartic form used in Eq.~\eqref{GL_3}). The mean-field solution corresponding to this configuration of the order parameters is displayed in Fig.~\ref{fig:hop2x2_SoD}d, showing the onset of a chiral nematic state with TrH order. In this case, the origin of nematicity is related to the different amplitudes of the order parameters rather than to the different phase, as described for the potential in Eq.~\eqref{GL_2}. Although only $\Delta_2$ has a finite and imaginary value, a finite current is observed along all three independent directions of the kagome lattice.

Considering now the configurations shown in Figs.~\ref{fig:hop2x2_SoD}b-c, we find a different GL potential:
\begin{align} \label{GL_4}
	\mathcal{F}_\text{SoD}^\text{b} & = \sum_j \psi_j^2 \big[ \alpha_1 + \alpha_2 \cos (2 \phi_j) + \frac{\beta}{4} \psi_j^2 \big] \nonumber \\
	& + \alpha \sum_{j, j' > j} \psi_j \psi_{j'} \cos (\phi_j) \cos (\phi_{j'}) .
\end{align}
Eq.~\eqref{GL_4} can be obtained from Eq.~\eqref{GL_3} by assuming $\alpha_3 = \alpha_4 = \alpha$. The quadratic term proportional to $\alpha$ is minimized, for $\alpha > 0$ ($\alpha < 0$), by $\phi_2 \ \text{mod} \ 2 \pi = \pi$, $\phi_{1,3} \ \text{mod} \ 2 \pi = 0$ ($\phi_j \ \text{mod} \ 2 \pi = 0$ or $\phi_j \ \text{mod} \ 2 \pi = \pi$) and $\psi_2 > \psi_1 = \psi_3$ ($\psi_1 = \psi_2 = \psi_3$); in any case, it tends to produce a real valued solution that does not break the TRS. However, a subtle interplay between $\alpha_2$ and $\alpha$ might lead to the stabilization of a state with $\psi_1 > \psi_2 = \psi_3$, $\phi_2 \ \text{mod} \ \pi = \pi/2$, $\phi_1 \ \text{mod} \ 2 \pi = 0$ and $\phi_3 \ \text{mod} \ 2 \pi = \pi$. Even in this case, the mean-field solutions show TrH features. Particularly, the ordered state depicted in Fig.~\ref{fig:hop2x2_SoD}f, besides breaking the TRS, shows nematicity. Recent x-ray diffraction measurements (combined with an unsupervised machine learning analysis) have found a TrH CBO qualitatively similar to the ones displayed in Figs.~\ref{fig:hop2x2}e-f and Figs.~\ref{fig:hop2x2_SoD}d-f \cite{Kautzsch2023_PRM}, even though these measurements are only sensitive to interatomic distances and not to the hopping strength.

%%%%%%% 3x3 UNIT CELL
\subsection{$\sqrt{3} \times \sqrt{3}$ unit cell}
In this section, we consider a 
%FG
CBO with 
%different ordering vector for the CBO on the kagome lattice, focusing on a 
$\sqrt{3} \times \sqrt{3}$ unit cell, represented in Figs.~\ref{fig:hopsq3xsq3}a-c. The patterns are characterized by symmetries of the CBO that do not imply any change of sign of the order parameters, leading to a larger number of non-zero contributions to the potential. The pattern shown in Fig.~\ref{fig:hopsq3xsq3}b leads to the GL free energy:
\begin{align} \label{GL_5}
	& \mathcal{F}_{\sqrt{3} \times \sqrt{3}}^\text{a} = \sum_j \psi_j \big[ h \cos (\phi_j) + \alpha_1 \psi_j + \alpha_2 \psi_j \cos (2 \phi_j) \big] \nonumber \\
	& + \alpha \sum_{j, j' > j} \psi_j \psi_{j'} \cos (\phi_j) \cos (\phi_{j'}) + \frac{8 \gamma}{3} \prod_j \psi_j \cos (\phi_j) \nonumber \\
	& + \frac{2}{3} \sum_{j,j' \neq j} \psi_j^2 \psi_{j'} \cos (\phi_{j'}) \big[ 2 \bar \gamma  \cos (2 \phi_j) + \gamma_5 \big] \nonumber \\
	& + \sum_j \psi_j^3 \big[ \frac{2 \gamma_6}{3} \cos (3 \phi_j) + 2 \gamma_7 \cos (\phi_j) + \frac{\beta}{4} \psi_j \big] ,
\end{align}
which reduces to Eq.~\eqref{GL_4} if $h=\gamma = \bar \gamma = \gamma_5 = \gamma_6 = \gamma_7 = 0$. To start with the analysis of Eq.~\eqref{GL_5}, we notice the presence of a linear contribution in the order parameters. For $h>0$ ($h<0$), this term is minimized by $\phi_j \ \text{mod} \ 2\pi = \pi$ ($\phi_j \ \text{mod} \ 2\pi = 0$). However, this term is not compatible with the zero order parameters we expect at high temperatures. For this reason, we are inclined to believe that $h$ must be zero for a faithful description of the CBO in kagome metals. The contribution related to $\bar \gamma$ is minimized, if $\bar \gamma < 0$ ($\bar \gamma > 0$), by $\phi_j \ \text{mod} \ 2 \pi = 0$ ($\phi_j  \ \text{mod} \ 2 \pi = \pi$) and $\psi_1 = \psi_2 = \psi_3$. We also analyze the cubic term proportional to $\gamma_5$; for $\gamma_5 > 0$ ($\gamma_5 < 0$), this is minimized by $\phi_j \ \text{mod} \ 2 \pi = \pi$ ($\phi_j \ \text{mod} \ 2 \pi = 0$) and $\psi_1 = \psi_2 = \psi_3$. Moreover, the term proportional to $\gamma_7$ stabilizes a solution $\phi_j \ \text{mod} \ 2 \pi = \pi$ when $\gamma_7 > 0$ and $\phi_j \ \text{mod} \ 2 \pi = 0$ when $\gamma_7 < 0$. Finally, the contribution proportional to $\gamma_6$ is minimized by $\phi_j \ \text{mod} \ 2 \pi/3 = \pi/3$ when $\gamma_6 > 0$ and by $\phi_j \ \text{mod} \ 2 \pi/3 = 0$ if $\gamma_6 < 0$, condition that might lead to complex order parameters.

Despite most of the contributions appearing in Eq.~\eqref{GL_5} tending to stabilize a state with real order parameters, one can select a proper combination of the interactions that lead to a breaking of the TRS. By taking into account a combination of $\alpha_2 > 0$, $\alpha > 0$ and $\bar \gamma > 0$ (besides $\alpha_1$ and $\beta$), one might get a solution with $\phi_{1,3} \ \text{mod} \ 2 \pi = 0$ and $\phi_2 \ \text{mod} \ \pi \neq 0$ and with $\psi_2 > \psi_1 = \psi_3$ (other two degenerate solutions can be obtained by exchanging $2 \leftrightarrow 1$ and $2 \leftrightarrow 3$, reflecting once again the $\mathbb{Z}_3$ symmetry of the problem). Even in this case, one can perform a mean-field calculation starting from Eq.~\eqref{ham_renorm}, which shows the onset of a CDW besides the CBO. This occurs together with the appearance of orbital currents along all the bonds and not just in the direction which explicitly breaks the TRS, see Fig.~\ref{fig:hopsq3xsq3}e. Assuming $\phi_j \ \text{mod} \ \pi = 0$ and $\Delta_j$ to have 
%FG
on-site 
%a local 
character, Eq.~\eqref{GL_5} describes an $l=0$ CDW with $\sqrt{3} \times \sqrt{3}$ unit cell.

The pattern in Fig.~\ref{fig:hopsq3xsq3}c shows three complex conjugations under the action of the generating symmetries, implying the GL free energy:
\begin{align} \label{GL_6}
	& \mathcal{F}_{\sqrt{3} \times \sqrt{3}}^\text{b} = \sum_j \psi_j \big[ h \cos (\phi_j) + \alpha_1 \psi_j + \alpha_2 \psi_j \cos (2 \phi_j) \big] \nonumber \\
	& + \sum_{j, j' > j} \psi_j \psi_{j'} \big[ \alpha_3 \cos (\phi_j + \phi_{j'}) + \alpha_4 \cos (\phi_j - \phi_{j'}) \big] \nonumber \\
         & + \frac{2}{3} \psi_1 \psi_2 \psi_3 \Big[ (\gamma_1 - \gamma_2) \cos \big( \sum_j \phi_j \big) + 4 \gamma_2 \prod_j \cos (\phi_j) \Big] \nonumber \\
	& + \frac{2}{3} \sum_{j, j' \neq j} \psi_j^2 \psi_{j'} \big[ \gamma_3 \cos (2 \phi_j + \phi_{j'}) + \gamma_4 \cos (2 \phi_j - \phi_{j'}) \big] \nonumber \\
	& + \frac{2 \gamma_5}{3} \sum_{j, j' \neq j} \psi_j^2 \psi_{j'} \cos (\phi_{j'}) \nonumber \\
	& + \sum_j \psi_j^3 \big[ \frac{2 \gamma_6}{3} \cos (3 \phi_j) + 2 \gamma_7 \cos (\phi_j) + \frac{\beta}{4} \psi_j\big] .
\end{align}
The above expression might be reduced to Eq.~\eqref{GL_5} by assuming $\alpha_3 = \alpha_4 = \alpha$, $\gamma_1 = \gamma_2 = \gamma$, $\gamma_3 = \gamma_4 = \bar \gamma$. Considering the term proportional to $\gamma_3$, one readily realizes that, for $\gamma_3 < 0$, it is minimized when $\phi_j \ \text{mod} \ 2 \pi = 0$ and $\psi_1 = \psi_2 = \psi_3$, while for $\gamma_3 > 0$ the minimum corresponds to $(\phi_1, \phi_2, \phi_3) = (\pi, \pi, \pi) \ \text{mod} \ 2 \pi$ and $\psi_1 = \psi_2 = \psi_3$. A similar result is found for $\gamma_4$. The several cubic contributions to the free energy in Eqs.~\eqref{GL_5} and \eqref{GL_6} might explain the strong first-order character of the CO transition observed in SbV$_6$Sn$_6$. When $\gamma_4 > 0$ appears together with $\alpha_2 > 0$, the competition between these two terms might stabilize a state with $\phi_2 \neq \phi_{1,3}$ and $\psi_2 > \psi_{1,3}$, leading to three complex values for the order parameters. The corresponding CDW and CBO break the mirror symmetry but preserve inversion symmetry. However, the current pattern breaks mirror and inversion symmetry, see Fig.~\ref{fig:hopsq3xsq3}f.

\begin{table*}
  \centering
  \begin{tabular}{ | C{2.2cm} | C{0.8cm} | C{0.8cm} | C{0.8cm} | C{0.8cm} | C{0.8cm} | C{1.3cm} | C{1.3cm} |}
    \hline
    $ $ & $\mathcal{F}_\text{Hex}$ & $\mathcal{F}_\text{TrH}^\text{a}$ & $\mathcal{F}_\text{TrH}^\text{b}$ & $\mathcal{F}_\text{SoD}^\text{a}$ & $\mathcal{F}_\text{SoD}^\text{b}$ & $\mathcal{F}_{\sqrt{3} \times \sqrt{3}}^\text{a}$ & $\mathcal{F}_{\sqrt{3} \times \sqrt{3}}^\text{b}$ \\ \hline
    Nematicity $\phi$ & \xmark & \xmark & \cmark & \cmark & \cmark & \cmark & \cmark \\ \hline
    Nematicity $\psi$ & \xmark & \xmark & \xmark & \cmark & \cmark & \cmark & \cmark \\
    \hline
  \end{tabular}
    \caption{For each of the potentials analyzed in the Sec.~\ref{sec:GL+MF}, the table shows if they can produce (\cmark) or not (\xmark) a nematic solution related to a phase ($\phi$) or to an amplitude ($\psi$) difference among the three order parameters.} \label{table:nem}
\end{table*}

\begin{table*}
  \centering
  \begin{tabular}{ | C{0.5cm} | C{0.5cm} | C{0.5cm} | C{0.5cm} | C{0.8cm} | C{0.5cm} | C{0.5cm} | C{0.8cm} | C{0.5cm} | C{0.5cm} | C{0.6cm} | C{0.6cm} | C{1.2cm} |}
    \hline
    $ $ & \multicolumn{3}{c|}{        $\mathcal{F}_\text{Hex}$        } & \multicolumn{1}{c|}{$\mathcal{F}_\text{TrH}^\text{a}$} & \multicolumn{2}{c|}{$\mathcal{F}_\text{TrH}^\text{b}$} & \multicolumn{1}{c|}{$\mathcal{F}_\text{SoD}^\text{a}$} & \multicolumn{2}{c|}{$\mathcal{F}_\text{SoD}^\text{b}$} & \multicolumn{2}{c|}{$\mathcal{F}_{\sqrt{3} \times \sqrt{3}}^\text{a}$} & \multicolumn{1}{c|}{$\mathcal{F}_{\sqrt{3} \times \sqrt{3}}^\text{b}$} \\ 
    $ $ & $1$d & $1$e & $1$f & $2$d & $2$e & $2$f & $3$d & $3$e & $3$f & $4$d & $4$e & $4$f \\ \hline
    $\sigma$ & \xmark & \cmark & \xmark & \xmark & \cmark & \xmark & \xmark & \xmark & \xmark & \cmark & \xmark & \xmark \\ \hline
    C$_3$ & \cmark & \cmark & \cmark & \cmark & \xmark & \xmark & \xmark & \xmark & \xmark & \cmark & \xmark & \xmark \\ \hline
    C$_2$ & \cmark & \xmark & \cmark & \cmark & \xmark & \cmark & \cmark & \xmark & \cmark & \cmark & \cmark & \xmark \\ 
    \hline
  \end{tabular}
    \caption{For each of the mean-field solutions analyzed in the Sec.~\ref{sec:GL+MF} and represented in Figs.1-4d-f, the table shows if they break (\xmark) or not (\cmark) the mirror symmetry ($\sigma$), the threefold rotation (C$_3$) and the inversion symmetry (C$_2$). 
    %FG
    A part from Fig.~\ref{fig:hopsq3xsq3}d, all 
    %All 
    the solutions break TRS. The corresponding free energy potential is displayed in the upper row.} \label{table:sym}
\end{table*}

%%%%%%% NEMATICITY
\subsection{Nematicity from the Ginzburg-Landau potentials} \label{subsec:nematicity}
Before we conclude this section, a few considerations are necessary. Immediately below the critical temperature $T_\text{C}$, for a second or higher order phase transition, the order parameters are infinitesimally small. As a consequence, for such a condition just the lowest contributions to the Ginzburg-Landau potential are relevant, i.e., the quadratic ones analyzed before (as we mentioned, we expect the linear terms in Eqs.~\eqref{GL_5}-\eqref{GL_6} to be zero). Since the nematic character of the solution of Eq.~\eqref{GL_2} is carried by the cubic term, we cannot expect this potential to describe the onset of nematicity at the phase transition. At the critical temperature of a first-order phase transition, instead, the above argument seems not applicable anymore, since the order parameters do not become infinitesimally small but discontinuously jump to a finite value. Since kagome metals $A$V$_3$Sb$_5$ ($A=$K, Rb, Cs) show a weakly first-order transition to the CO phase, the considerations we made for a second-order transition should remain substantially valid for this class of compounds. Given the presence of the cubic interactions, the potential Eq.~\eqref{GL_2} can describe both the onset of nematicity at and below the critical temperature depending by the choice of the GL parameters.

On the other hand, Eqs.~\eqref{GL_3}-\eqref{GL_4} do not have any cubic terms, thus the transition described by these potentials has to be (at least) second order. 
%FG
The nematicity is driven, in these cases, by the interplay among several quadratic contributions to the free energy. To have a nematic CO at $T_\text{C}$ electronic nematicity has to be developed already in the high-temperature metal. 
%The nematicity is driven, in these cases, by the interplay among several quadratic contributions to the free-energy. Since the solution at the phase transition must inherit the symmetry properties of the higher temperature state, to have a nematic CO at $T_\text{C}$ electronic nematicity has to be developed already in the high-temperature metal. 
If, instead, the metal has sixfold rotational symmetry (as we are assuming), this implies that $\alpha_3$, $\alpha_4$ and $\alpha$ in Eqs.~\eqref{GL_3}-\eqref{GL_4} must satisfy some constrains at the transition point, e.g. they might be very small (in absolute value) compared to the other quadratic interactions. Nevertheless, this constraint does not have to be satisfied at lower temperatures, implying that nematicity can still develop below $T_\text{C}$. In the case nematicity already takes place in the high temperature metallic state, the coefficients of the GL potentials become dependent by the index of the order parameters $\Delta_1$, $\Delta_2$ and $\Delta_3$. To provide an example, the quadratic term proportional to $\alpha_1$ is replaced by $\alpha_1 \sum_j \psi_j^2 \rightarrow \sum_j \alpha_{1,j} \psi_j^2$ and analogously for the other contributions to the free energy.

Similar considerations as outlined in the previous two paragraphs apply also for the potentials in Eqs.~\eqref{GL_5}-\eqref{GL_6}.

From the previous analysis, we conclude that nematicity might arise in kagome metals in several ways: from a phase difference among the three order parameters, and from a combination of different phases and different amplitudes. Particularly, the first case seems to be supported by the GL potential in Eq.~\eqref{GL_2}, corresponding to a TrH ordering; the second scenario is instead supported by the potentials in Eqs.~\eqref{GL_3}-\eqref{GL_4} (corresponding to a SoD distortion) and by Eqs.~\eqref{GL_5}-\eqref{GL_6} (for the $\sqrt{3} \times \sqrt{3}$ unit cell). As we analyze in the next section, RUS might help distinguish between these cases and it might provide additional information on the critical temperature for the onset of nematicity.
%FG
We further stress that, within this analysis, we are not considering the possibility for the order parameters $\Delta_1$, $\Delta_2$ and $\Delta_3$ to have an out-of-plane component, which might explain nematicity due to a phase-shift of the CBO in consecutive kagome layers. The key results obtained in this section are summarized in Tables \ref{table:nem}-\ref{table:sym}.
%

%%%%%%%%%%%%%%%%%%%%%%%%%%%%%%%%%%%%%%%%%%%%%%% RUS
\section{Resonant ultrasound spectroscopy} \label{sec:RUS}
Resonant ultrasound spectroscopy measures the discontinuities in the elements of the stiffness tensor at the critical temperature, which can be related to the symmetry properties of the order parameters at the phase transition. In this section, we aim to develop a theoretical description of RUS for the CO in kagome metals. To do that, we consider the 
%FG
free energy 
%free-energy 
contributions for the elastic deformations of the solid $\epsilon$ and the coupling between the order parameters $\Delta_j$ and $\epsilon$. The space group of the kagome metals is P$6/$mmm, with point group D$_{6 \text{h}}$. Given the quadratic representations of this point group, we might decompose the deformation tensor $\epsilon_{pq} = \frac{1}{2} (\partial_p u_q + \partial_q u_p)$, where $u_q$ is the q-th component of the local deformation vector ($p,q=x,y,z$), into the irreducible representations (irreps) $\epsilon_{\text{A}_{1 \text{g},1}} = \epsilon_{xx} + \epsilon_{yy}$, $\epsilon_{\text{A}_{1 \text{g},2}} = \epsilon_{zz}$, $\epsilon_{\text{E}_{1 \text{g}}} = (2 \epsilon_{xz}, 2 \epsilon_{yz})$ and $\epsilon_{\text{E}_{2 \text{g}}} = (\epsilon_{xx} - \epsilon_{yy}, 2 \epsilon_{xy})$ \cite{Nie2022_Nat}. Here, we choose the coordinate system so that the $x$ and the $y$ axis belong to the plane containing the kagome lattice formed by the vanadium atoms, while the $z$ axis is orthogonal to this plane. The $\epsilon_{\text{A}_{1 \text{g},1}}$ and $\epsilon_{\text{A}_{1 \text{g},2}}$ deformations change the volume of the system and, for this reason, are called compressional, while $\epsilon_{\text{E}_{1 \text{g}}}$ and $\epsilon_{\text{E}_{2 \text{g}}}$ are called shear deformations because they preserve the total volume even if they break the hexagonal symmetry of the lattice. By moving to the Voigt notation, which maps the six independent components of the deformation tensor into a vector $\boldsymbol{\epsilon} = (\epsilon_1, \epsilon_2, \epsilon_3, \epsilon_4, \epsilon_5, \epsilon_6) = (\epsilon_{xx}, \epsilon_{yy}, \epsilon_{zz}, 2 \epsilon_{yz}, 2 \epsilon_{xz}, 2 \epsilon_{xy})$, one can write the elastic (el) contribution to the free energy as:
\begin{align}
    \mathcal{F}_\text{el} = \frac{1}{2} \sum_{i,k = 1}^{6} \epsilon_i c_{ik} \epsilon_k ,
\end{align}
with $c_{ik}$ the stiffness matrix, which, for a system with D$_{6 \text{h}}$ symmetry, has only five independent components \cite{LandauBook7}:
\begin{align} \label{stiff_mat}
    c = \begin{pmatrix}
    c_{11} & c_{12} & c_{13} & 0 & 0 & 0 \\
    c_{12} & c_{11} & c_{13} & 0 & 0 & 0 \\
    c_{13} & c_{13} & c_{33} & 0 & 0 & 0 \\
    0 & 0 & 0 & c_{44} & 0 & 0 \\
    0 & 0 & 0 & 0 & c_{44} & 0 \\
    0 & 0 & 0 & 0 & 0 & c_{66}
    \end{pmatrix} ,
\end{align}
since $c_{66} = \frac{c_{11} - c_{12}}{2}$. The allowed contributions appearing in the free energy must couple terms in the electronic order parameter and in the elastic deformations which share the same symmetry. We consider the order parameters to belong to the two single-component A$_{1 \text{g}}$ or to the two-components E$_{2 \text{g}}$ irreps. In the following, we analyze both these cases. 

%%%%%% ONE-COMPONENT
\subsection{One-component order parameters}
%FG
We consider the three order parameters $\Delta_j$ to have one component. 
%The three $\Delta_j$ can be one-component order parameters, corresponding to the A$_{1 \text{g}}$ irrep of the point group D$_{6 \text{h}}$. 
Particularly, 
%FG
on-site 
%local 
and real $\Delta_j$ would describe a CDW \cite{McMillan1975_PRB,vanWezel2011_EPL,Denner2021_PRL}. The ordered state breaks at least the translational symmetry, thus the lowest-order contributions to the interaction (int) 
%FG
free energy 
%free-energy 
between the elastic deformations and the order parameters is:
\begin{align} \label{GL_int_psi_1}
    & \mathcal{F}_\text{int} = (g_1 \epsilon_{\text{A}_{1 \text{g},1}} + g_2 \epsilon_{\text{A}_{1 \text{g},2}}) \sum_j \psi_{j}^2 ,
\end{align}
where $g_1$ ($g_2$) is the coupling constant of the order parameter with the $\text{A}_{1 \text{g},1}$ ($\text{A}_{1 \text{g},2}$) irrep of the elastic tensor. The total (tot) potential of the problem reads:
\begin{align}
    \mathcal{F}_\text{tot} = \mathcal{F} + \mathcal{F}_\text{int} + \mathcal{F}_\text{el} ,
\end{align}
with $\mathcal{F}$ the contribution to the free energy coming from the order parameters, which, for a CDW with real $\Delta_j$, correspond to Eq.~\eqref{GL_1} or to Eq.~\eqref{GL_5} with the prescriptions commented above. At the phase transition, when $\Delta_j$ starts to become finite, we expect a sudden coupling with the elastic deformations due to Eq.~\eqref{GL_int_psi_1}. This interaction produces a discontinuity in the components of the stiffness matrix at the critical temperature, which can be measured with RUS. By introducing the vector of the order parameters $\boldsymbol{\Phi} = (\psi_1, \psi_2, \psi_3, \phi_1, \phi_2, \phi_3)$, one can provide the expression for the discontinuity of the stiffness matrix elements as \cite{Slonczewski1970_PRB,Carpenter1998_EJM,Ghosh2021_Nat_Phys}:
\begin{align} \label{disc_gen}
    \Delta c_{ik} = \sum_{n,m = 1}^{6} \frac{\partial^2 \mathcal{F}_\text{int}}{\partial \epsilon_i \partial \Phi_n} \Big( \frac{\partial^2 \mathcal{F}}{\partial \boldsymbol{\Phi}^2} \Big)^{-1}_{n,m} \frac{\partial^2 \mathcal{F}_\text{int}}{\partial \epsilon_k \partial \Phi_m} ,
\end{align}
where $\frac{\partial^2 \mathcal{F}}{\partial \boldsymbol{\Phi}^2}$ is the Hessian of the GL potential.

If a nematic CO is stabilized below $T_\text{C}$ ($T_\text{nem} < T_\text{C}$), the discontinuities of the stiffness matrix at the onset of the translation symmetry breaking ($T_\text{C}$) are:
\begin{align} \label{disc_1}
    \Delta c_{11} = \Delta c_{22}, \ \Delta c_{33}, \ \Delta c_{12}, \ \Delta c_{13} = \Delta c_{23} ,
\end{align}
together with the consistency relation $\frac{\Delta c_{11} + \Delta c_{12}}{2} \Delta c_{33} = (\Delta c_{13})^2$. Since Eq.~\eqref{GL_int_psi_1} is insensible to the phases of the order parameters $\phi_1$, $\phi_2$ and $\phi_3$ and to the relative difference in the amplitudes $\psi_1$, $\psi_2$ and $\psi_3$, the onset of the nematic order at $T_\text{nem} < T_\text{C}$ would not be marked by a finite value of $\Delta c_{ik}$.

As we have already mentioned, to have $T_\text{nem} = T_\text{C}$, the system must show nematicity already in the higher temperature metallic state. In that case, the starting stiffness matrix would not be Eq.~\eqref{stiff_mat} but rather the one of a system with point group C$_2$, i.e., with thirteen independent components. Also the Ginzburg-Landau 
%FG
free energy 
%free-energy 
for the order parameters has to be changed as described in Sec.~\ref{subsec:nematicity}. Similarly, the lowest order contributions to the interaction part of the free energy Eq.~\eqref{GL_int_psi_1} has to be modified accordingly:
\begin{align} \label{GL_int_psi_1_C2}
    & \mathcal{F}_\text{int}^{\text{C}_2} = (g_1 \epsilon_{\text{A}_1} + g_2 \epsilon_{\text{A}_2} + g_3 \epsilon_{\text{A}_3} + g_4 \epsilon_{\text{A}_4}) \sum_j \psi_{j}^2 ,
\end{align}
where $\epsilon_{\text{A}_1} = \epsilon_{xx}$, $\epsilon_{\text{A}_2} = \epsilon_{yy}$, $\epsilon_{\text{A}_3} = \epsilon_{zz}$ and $\epsilon_{\text{A}_4} = \epsilon_{xy}$ are irreps of C$_2$. In this case, the discontinuities of the stiffness matrix at $T_\text{C}$ are:
\begin{align} \label{disc_1_neq}
    & \Delta c_{11} \neq \Delta c_{22}, \ \Delta c_{33}, \ \Delta c_{66}, \ \Delta c_{12}, \ \Delta c_{13} \neq \Delta c_{23} , \nonumber \\
    & \Delta c_{16}, \ \Delta c_{26}, \ \Delta c_{36} .
\end{align}

%%%%%% TWO-COMPONENTS
\subsection{Two-component order parameters}
%FG
We now consider the order parameters $\Delta_j$ to have a two-components representation. 
%We now consider the order parameters $\Delta_j$ to have a two-components representation, which correspond to the E$_{2 \text{g}}$ irrep of the point group D$_{6 \text{h}}$. 
Since, for the CBO, each $\Delta_j$ has both an amplitude and a direction (the latter is provided by the spatial orientation of each $\Delta_j$, as represented, e.g., in Figs.~\ref{fig:hop2x2}a-c), they can be regarded as two-dimensional vectors $\boldsymbol{\Delta}_j = (\Delta_{j,x}, \Delta_{j,y}) = \Delta_j \big(\cos (\theta_j), \sin (\theta_j) \big)$ with amplitude $\Delta_j = \sqrt{\Delta_{j,x}^2 + \Delta_{j,y}^2}$. If no strain is applied to the system, the angles $\theta_j$ are fixed and we can write $\boldsymbol{\Delta}_1 = \frac{\Delta_1}{2} (1, -\sqrt{3})$, $\boldsymbol{\Delta}_2 = \Delta_2 (1, 0)$ and $\boldsymbol{\Delta}_3 = \frac{\Delta_3}{2} (1, \sqrt{3})$, having assumed a frame of reference with the x-axis parallel to $\boldsymbol{\Delta}_2$. We might write the interaction with the elastic deformations to the lowest order in $\Delta$ and $\epsilon$ as \cite{Sigrist2002_PTP}:
\begin{align} \label{GL_int_psi}
    \mathcal{F}_\text{int} & = (g_1 \epsilon_{\text{A}_{1 \text{g},1}} + g_2 \epsilon_{\text{A}_{1 \text{g},2}}) \sum_j \psi_{j}^2 \nonumber \\
    & + g_3 \epsilon_{\text{E}_{2 \text{g},1}} \Big( \psi_2^2 - \frac{\psi_1^2 + \psi_3^2}{2} \Big) \nonumber \\
    & + \frac{\sqrt{3}}{2} g_3 \epsilon_{\text{E}_{2 \text{g},2}} (\psi_3^2 - \psi_1^2) ,
\end{align}
where the coupling constant $g_3$ to $\epsilon_{\text{E}_{2 \text{g},1}}$ and $\epsilon_{\text{E}_{2 \text{g},2}}$ is the same because they belong to the same irrep $\text{E}_{2 \text{g}}$. In principle, a coupling of the order parameters with $\epsilon_{\text{E}_{1 \text{g}}}$ of the kind $\epsilon_{\text{E}_{1 \text{g}}}^2 \sum_j \psi_{j}^2$ would be allowed by symmetry. However, this contribution is higher-order with respect to the other terms in Eq.~\eqref{GL_int_psi} and is not expected to provide any discontinuity in any component of the stiffness matrix; at most, it would provide a change of slope in $c_{44}$ at the phase transition \cite{Theuss2022_PRB}. Given the more complex structure of the interaction between the order parameters and the elastic deformations with respect to the single-component case (compare Eq.~\eqref{GL_int_psi_1} and Eq.~\eqref{GL_int_psi}), we expect not only a different functional dependence of the discontinuities of the stiffness matrix components, but also a qualitative difference. In the case in which the ordered state reached at $T_\text{C}$ is characterized by order parameters with the same amplitude $\psi_1 = \psi_2 = \psi_3$, we obtain:
\begin{align} \label{disc_2_eq}
    & \Delta c_{11} = \Delta c_{22} , \ \Delta c_{33} , \ \Delta c_{66}, \ \Delta c_{12} , \ \Delta c_{13} = \Delta c_{23} ,
\end{align}
with the same consistency relation discussed below Eq.~\eqref{disc_1}. Differently from a one-component order parameter, in this case a discontinuity in $c_{66}$ is expected to be finite.

Below the critical temperature for the onset of the CBO, the system might go through the nematic transition, moving from a state with $\psi_1 = \psi_2 = \psi_3 (= \psi)$ to one with $\psi_2 \neq \psi_1 = \psi_3$. By assuming this transition to be of the second order, one might compute the discontinuities of the stiffness matrix using Eq.~\eqref{disc_gen} starting from $\psi \neq 0$. In this equation, the same expression of $\mathcal{F}_\text{int}$ considered above can be used, i.e., Eq.~\eqref{GL_int_psi}. However, the free energy $\mathcal{F}$ is not just the GL potential for the order parameters as it was in the previous cases and it also has the contribution coming from the finite value of the deformations $\epsilon$. With respect to Eq.~\eqref{disc_2_eq}, this produce the additional differentiation:
\begin{align} \label{disc_2_neq}
    & \Delta c_{11} \neq \Delta c_{22}, \ \Delta c_{13} \neq \Delta c_{23} .
\end{align}
Since the point group C$_2$ has only one-component irreps, it does not make sense to discuss the case in which nematicity occurs at $T_\text{C}$ when the order parameters have two components.

%%%%%%%%%%%%%%%%%%%%%%%%%%%%%%%%%%%%%%%%%%%%%%% DISC+CONCL
\section{Discussion and Conclusions} \label{sec:Conclusions}
%FG
In this work, we have developed a general Ginzburg-Landau theory for the charge order observed in kagome metals based on 
%Given the contradictory experimental and theoretical findings regarding the origin and the form of the charge order in kagome metals, we have developed a general Ginzburg-Landau theory for this state based on 
the assumptions of having a $3$Q ordering, an in-plane $2 \times 2$ or $\sqrt{3} \times \sqrt{3}$ reconstruction and, in most of the cases, a high-temperature metal with the same point group symmetries of the kagome lattice. Our mean-field analysis shows that different $2 \times 2$ patterns, such as the tri-hexagonal or the star of David ones, can induce a charge order compatible with the experimental indications, i.e., a state with broken translation and time-reversal symmetries with nematic character, see panel f of Fig.~\ref{fig:hop2x2} and \ref{fig:hop2x2_SoD}. The corresponding Ginzburg-Landau potentials are presented in Eq.~\eqref{GL_2} and in Eq.~\eqref{GL_4}, respectively. Instead, the real order parameter limit of Eqs.~\eqref{GL_5}-\eqref{GL_6}, which produces the $\sqrt{3} \times \sqrt{3}$ charge order shown in Fig.~\ref{fig:hopsq3xsq3}d, seems compatible with experiments on ScV$_6$Sn$_6$ \cite{Arachchige2022_PRL}. The analysis we have performed might be relevant also for other kagome metals that might be discovered in the future \cite{Jiang2022_ChinPhysLett,Yi2022_PRB,Yang2022_arXiv_CTB,Li2022_arXiv2}.

Concerning the order parameters, there are no clear indications regarding their number of components. Moreover, different experiments do not agree on the temperature $T_\text{nem}$ for the transition to the nematic charge order, with some suggesting that this state is reached below the critical temperature for the charge ordering $T_\text{C}$, and others indicating that they occur together ($T_\text{nem} = T_\text{C}$). We suggest resonant ultrasound spectroscopy as an experimental tool to clarify these aspects. Indeed, our symmetry analysis implies that if the nematic charge order is stabilized at the critical temperature $T_\text{C}$, the order parameters must have one component. The corresponding discontinuities in the components of the stiffness matrix at the transition are reported in Eq.~\eqref{disc_1_neq}. 

Now, suppose the nematic state with C$_2$ symmetry is stabilized at lower temperatures than the critical temperature for the charge bond order phase transition $T_\text{nem} < T_\text{C}$. Then, if the order parameters have a single component each, we should not expect a jump in the $66$ component of the stiffness matrix at $T_\text{C}$ ($\Delta c_{66} = 0$). In contrast, the opposite has to occur if the order parameters have two components ($\Delta c_{66} \neq 0$), compare Eq.~\eqref{disc_1} and Eq.~\eqref{disc_2_eq}. At $T_\text{nem}$, instead, if nematicity is due to a difference in the amplitudes and not only in the phases of $\Delta_1$, $\Delta_2$ and $\Delta_3$, one would observe another set of discontinuities in the elements of the stiffness matrix at this lower temperature only in case that the order parameters have two components, see Eq.~\eqref{disc_2_neq}. In this case, the Ginzburg-Landau potential in Eq.~\eqref{GL_4} is the preferred candidate for describing the $2 \times 2$ charge order of kagome metals. Indeed, this potential allows the three amplitudes of the order parameters to become different at $T_\text{nem} < T_\text{C}$.

Once the properties of the ground state are finally clarified by experiments such as the one we are proposing, it would be of interest to study the enhancement of nematicity, or to select one of the states related by the emergent 
%FG
$\mathbb{Z}_3$ 
%$\mathbf{Z}_3$ 
symmetry by applying a finite strain to the system. A similar procedure has recently shown its power in controlling the anomalous Hall effect in the Weyl antiferromagnet Mn$_3$Sn \cite{Ikhlas2022_NatPhys,Dasgupta2022_PRB}. Another fruitful avenue for future research is the application of short laser pulses to kagome metals to study their behavior under nonthermal conditions \cite{delaTorre2021,Azoury2023_arXiv,Ratcliff2021_PRM}, which might give rise to the opportunity of polarization-selective control over the multiple order parameter components, similar to the case of multi-component superconductors \cite{claassen_universal_2019}. Other interesting avenues are to study the interplay among the $2 \times 2$ charge order and in-plane ferromagnetism observed in the iron-based kagome layers of FeGe \cite{Teng2022_Nat,Yin2022_PRL} with Ginzburg-Landau theories, or to investigate the role of a near-Fermi level flat band as observed in the kagome compound Ni$_3$In \cite{Ye2021_arXiv}.

\begin{acknowledgements}
F.G. acknowledges stimulating discussions with Francesco Ferrari, Gregorio de la Fuente Simarro, Jonas Hauck, Lennart Klebl, M. Michael Denner, Giacomo Passetti and Brad Ramshaw. Simulations were performed with computing resources granted by RWTH Aachen University under project rwth1230. 
F.G. and D.M.K. acknowledge support by the DFG via Germany’s Excellence Strategy$-$Cluster of Excellence Matter and Light for Quantum Computing (ML$4$Q, Project No. EXC $2004/1$, Grant No. $390534769$), within the RTG 1995 and within the Priority Program SPP 2244 ``2DMP''. 
A.C. and R.T. acknowledge support from the DFG through QUAST FOR $5249-449872909$ (Project P$3$), through Project-ID $258499086-$SFB $1170$, and from the W\"urzburg-Dresden Cluster of Excellence on Complexity and Topology in Quantum Matter$-$ct.qmat Project$-$ID $390858490-$EXC $2147$. 
M.A.S. acknowledges financial support through the Deutsche Forschungsgemeinschaft (DFG, German Research Foundation) via the Emmy Noether program (SE $2558/2$).
\end{acknowledgements}

%\bibliography{sources.bib}

\newpage~\newpage~

\appendix 
\onecolumngrid

\section*{Supplemental Material for: 'Theory of nematic charge orders in kagome metals'}

In this Supplemental Material, we present some additional information concerning the quartic interaction of the Ginzburg-Landau potentials and on the mean-field calculation discussed in the main text. 

%%%%%%%%%%%%%%%%%%%%%%%%%%%%%
\subsection{Quartic interaction of the Ginzburg-Landau potential}
The full expression for the quartic part of the Ginzburg-Landau potentials in Eqs.~\eqref{GL_0},\eqref{GL_1}-\eqref{GL_2} of the main text reads:
\begin{align} \label{GL_0_quart}
    \mathcal{F}_{\text{quart}} & = \frac{\beta_1}{4} \sum_j \psi_j^4 + \frac{\beta_2}{2} \sum_j \psi_j^4 \cos (4 \phi_j) + \frac{\beta_3}{4} \sum_{j,j'>j} \psi_j^2 \psi_{j'}^2 + \frac{\beta_4}{2} \sum_j \psi_j^4 \cos (2 \phi_j) \nonumber \\
    & + \frac{\beta_5}{2} \sum_{j, j'>j} \psi_j^2 \psi_{j'}^2 \cos [2 (\phi_j + \phi_{j'}) ] + \frac{\beta_6}{2} \sum_{j,j' > j} \psi_j^2 \psi_{j'}^2 \cos [2 (\phi_j - \phi_{j'}) ] + \frac{\beta_7}{2} \sum_{j, j' \neq j} \psi_j^2 \psi_{j'}^2 \cos (2 \phi_{j'}) .
\end{align}
The quartic potential that appears, e.g., in Eq.~\eqref{GL_0} of the main text can be obtained from Eq.~\eqref{GL_0_quart} assuming $\beta_1 = \beta$, $\beta_2 = \beta_3 = \beta_4 = \beta_5 = \beta_6 = \beta_7 = 0$. This way, it is possible to obtain an analytical expression for the solutions of Eq.~\eqref{GL_0} of the main text. One can easily derive the full expressions for the potentials in Eqs.~\eqref{GL_3}-\eqref{GL_6} of the main text, even if they are more involved. For brevity, we do not report them here.

%%%%%%%%%%%%%%%%%%%%%%%%%%%%%
\subsection{Mean-field analysis}
We study the Hamiltonian Eq.~\eqref{ham} of the main text on the kagome lattice in the presence of the field patterns shown in Figs.~\ref{fig:hop2x2_Fe}-\ref{fig:hopsq3xsq3}. We perform a mean-field decoupling of the interaction \cite{Wen2010_PRB,Liu2010_PRB,Lopez2020_PRB}:
\begin{align}
    & n_{i, \uparrow} n_{i, \downarrow} \approx \langle n_{i, \uparrow} \rangle  n_{i, \downarrow} + \langle n_{i, \downarrow} \rangle  n_{i, \uparrow} - \langle n_{i, \uparrow} \rangle \langle n_{i, \downarrow} \rangle \;, \\
    & n_{i} n_{j} \approx \langle n_{i} \rangle n_{j} + \langle n_{j} \rangle n_{i} - \langle n_{i} \rangle \langle n_{j} \rangle - \sum_{\sigma} \big( \langle c^\dagger_{i,\sigma} c_{j,\sigma} \rangle c^\dagger_{j,\sigma} c_{i,\sigma} + \langle c^\dagger_{j,\sigma} c_{i,\sigma} \rangle c^\dagger_{i,\sigma} c_{j,\sigma} - \langle c^\dagger_{i,\sigma} c_{j,\sigma} \rangle \langle c^\dagger_{j,\sigma} c_{i,\sigma} \rangle \big) ,
\end{align}
having assumed the symmetry breaking to occur along the spin z direction. We consider two possibilities for the unit cell of the problem depending by the specific pattern for the CBO we analyze, one with $9$ atoms ($\sqrt{3} \times \sqrt{3}$ unit cell) and one with $12$ atoms ($2 \times 2$ unit cell). The number of variation parameters is $54$ (of which $36$ are complex) in the former and $72$ (of which $48$ are complex) in the latter case.

After the mean-field decoupling, the Hamiltonian of the problem becomes quadratic, thus it can be easily diagonalized in reciprocal space at each $\mathbf{k}$ point. Starting from an initial guess for the variational parameters, we can write:
\begin{align}
    H^{\text{MF}} = \sum_{\mathbf{k},\sigma} \psi^\dagger_{\mathbf{k},\sigma} \mathcal{H}_{\mathbf{k},\sigma} [ \langle n_{i,\sigma '} \rangle, \langle c^\dagger_{i,\sigma'} c_{j,\sigma'} \rangle ] \psi_{\mathbf{k},\sigma} = \sum_{\mathbf{k},\sigma} \phi^\dagger_{\mathbf{k},\sigma} \mathcal{H}^d_{\mathbf{k},\sigma} [ \langle n_{i,\sigma '} \rangle, \langle c^\dagger_{i,\sigma'} c_{j,\sigma'} \rangle ] \phi_{\mathbf{k},\sigma} ,
\end{align}
where $\mathcal{H}_{\mathbf{k},\sigma} [ \langle n_{i,\sigma '} \rangle, \langle c^\dagger_{i,\sigma'} c_{j,\sigma'} \rangle ]$ is the Bloch Hamiltonian of the problem that depends by all the variational parameters $\langle n_{i,\sigma '} \rangle$ and $\langle c^\dagger_{i,\sigma'} c_{j,\sigma'} \rangle$, while $\psi_{\mathbf{k},\sigma}$ is the reciprocal space spinor with the dimension of the unit cell containing the annihilation operators. $\mathcal{H}^d_{\mathbf{k},\sigma} [ \langle n_{i,\sigma '} \rangle, \langle c^\dagger_{i,\sigma'} c_{j,\sigma'} \rangle ]$ is the diagonal form of the Bloch Hamiltonian with eigenvalues $E_{\mathbf{k},\sigma, m}$ on the diagonal and
\begin{align}
    \phi_{\mathbf{k},\sigma} = U_{\mathbf{k},\sigma} [ \langle n_{i,\sigma '} \rangle, \langle c^\dagger_{i,\sigma'} c_{j,\sigma'} \rangle ] \psi_{\mathbf{k},\sigma} ,
\end{align}
is the spinor containing the eigenoperators at point $\mathbf{k}$ and spin $\sigma$, with $U_{\mathbf{k},\sigma}$ the unitary transformation that transforms $\psi_{\mathbf{k},\sigma}$ into $\phi_{\mathbf{k},\sigma}$ (here and in the following, we omit the functional dependence of $U_{\mathbf{k},\sigma}$ by the variational parameters for conciseness). We compute the expectation value:
\begin{align} \label{exp_val}
    \langle c^\dagger_{i,\sigma} c_{j,\sigma} \rangle & = \frac{1}{A_{BZ}} \sum_{\mathbf{k}} e^{-i \mathbf{k} \cdot (\mathbf{R}_i - \mathbf{R}_j)} \langle c^\dagger_{i,\mathbf{k},\sigma} c_{j,\mathbf{k},\sigma} \rangle = \frac{1}{A_{BZ}} \sum_{\mathbf{k}} e^{-i \mathbf{k} \cdot (\mathbf{R}_i - \mathbf{R}_j)} \langle ( \psi^\dagger_{\mathbf{k},\sigma} )_i ( \psi_{\mathbf{k},\sigma} )_j \rangle \nonumber \\
    & = \frac{1}{A_{BZ}} \sum_{\mathbf{k}} e^{-i \mathbf{k} \cdot (\mathbf{R}_i - \mathbf{R}_j)} \sum_{m,l} \langle ( \phi^\dagger_{\mathbf{k},\sigma} )_m ( U_{\mathbf{k},\sigma} )_{m i} ( U^\dagger_{\mathbf{k},\sigma} )_{j l}  ( \phi_{\mathbf{k},\sigma} )_l \rangle \nonumber \\
    & = \frac{1}{A_{BZ}} \sum_{\mathbf{k}} e^{-i \mathbf{k} \cdot (\mathbf{R}_i - \mathbf{R}_j)} \sum_{m}( U_{\mathbf{k},\sigma} )_{m i} ( U^\dagger_{\mathbf{k},\sigma} )_{j m} f (E_{\mathbf{k},\sigma,m} - \mu) .
\end{align}
$\mathbf{R}_i$ and $\mathbf{R}_j$ are the real space positions of the sites $i$ and $j$, respectively, $A_{BZ}$ is the area of the Brillouin zone and $f(E_{\mathbf{k},\sigma,m} - \mu)$ is the Fermi distribution function computed at the eigenvalue of the problem shifted by the chemical potential $\mu$. $\mu$ is determined by fixing the number of particles on the unit cell to the desired value ($2.5$ electrons every three sites at the p-type van Hove singularity).

Eq.~(\ref{exp_val}) permits to compute a new value of the variational parameters given the initial guess. By iterating this procedure, we can reach a self consistent solution that generally depends by the original guess. This procedure is analogous to minimizing the free-energy:
\begin{align}
    F = - \frac{1}{A_{BZ} \beta} \sum_{\mathbf{k}, \sigma, m} \ln \big[ 1 + e^{- \beta (E_{\mathbf{k},\sigma,m} - \mu)} \big] + \mu N + F^{\text{MF}} ,
\end{align}
with $N$ the total number of electrons on the unit cell and $F^{\text{MF}}$ the mean-field free energy:
\begin{align}
    F^{\text{MF}} = - U \sum_{i} \langle n_{i, \uparrow} \rangle \langle n_{i, \downarrow} \rangle - V \sum_{\langle i,j \rangle} \langle n_i \rangle \langle n_j \rangle + V \sum_{\langle i,j \rangle, \sigma} \langle c^\dagger_{i,\sigma} c_{j,\sigma} \rangle \langle c^\dagger_{j,\sigma} c_{i,\sigma} \rangle ,
\end{align}
where the summation over the nearest neighbors has to count each bond once.

\end{document}